\documentclass{article}

\usepackage[english]{babel}

\usepackage[letterpaper,top=2cm,bottom=2cm,left=3cm,right=3cm,marginparwidth=1.75cm]{geometry}

\usepackage{csquotes}
\usepackage{hhline}
\usepackage[autocite = superscript, style=numeric,sorting=none,giveninits=true]{biblatex}
\DeclareNameAlias{author}{family-given}

\makeatletter
\renewcommand\@biblabel[1]{#1.}
\makeatother

\makeatletter
\renewbibmacro*{textcite}{%
  \iffieldequals{namehash}{\cbx@lasthash}
    {\mkbibsuperscript{\supercitedelim}}
    {\cbx@tempa
     \ifnameundef{labelname}
       {\printfield[citetitle]{labeltitle}}
       {\printnames{labelname}}}%
  \ifnumequal{\value{citecount}}{1}
    {}
    {}%
  \mkbibsuperscript{\usebibmacro{cite}}%
  \savefield{namehash}{\cbx@lasthash}%
  \gdef\cbx@tempa{\addspace\multicitedelim}}%

\DeclareCiteCommand{\textcite}
  {\let\cbx@tempa=\empty
   \undef\cbx@lasthash
   \iffieldundef{prenote}
     {}
     {\BibliographyWarning{Ignoring prenote argument}}%
   \iffieldundef{postnote}
     {}
     {\BibliographyWarning{Ignoring postnote argument}}}
  {\usebibmacro{citeindex}%
   \usebibmacro{textcite}}
  {}
  {}
\makeatother

\usepackage{amsmath,amsthm,amsfonts}

\usepackage{bbm}
\usepackage{graphicx}
\usepackage{algorithm}
\usepackage{algpseudocode}
\usepackage{soul}
\usepackage[colorlinks=true, allcolors=blue]{hyperref}
\usepackage{pgf,tikz}
\usepackage{cancel}
\usepackage[defaultcolor=red]{changes}
\usetikzlibrary{arrows,shapes.arrows,shapes.geometric,shapes.multipart,decorations.pathmorphing,positioning}

\newtheorem*{assumption*}{\assumptionnumber}
\providecommand{\assumptionnumber}{}
\makeatletter
\newenvironment{newassumption}[2]
{%
	\renewcommand{\assumptionnumber}{Assumption #1~{\normalfont~(#2)}}%
	\begin{assumption*}%
		\protected@edef\@currentlabel{#1}%
	}
	{%
	\end{assumption*}
}
\makeatother

\newtheorem*{result*}{\resultnumber}
\providecommand{\resultnumber}{}
\makeatletter
\newenvironment{newresult}[1]
{%
	\renewcommand{\resultnumber}{Result #1}%
	\begin{result*}%
		\protected@edef\@currentlabel{#1}%
	}
	{%
	\end{result*}
}
\makeatother

\def\indep{\!\perp\!\!\!\perp}

\newtheorem{assumption}{Assumption}
\newtheorem{result}{Result}
\title{Regression-based proximal causal inference for right-censored time-to-event data}
\date{}

\begin{document}

\begin{titlepage} % Suppresses displaying the page number on the title page and the subsequent page counts as page 1
	\newcommand{\HRule}{\rule{\linewidth}{0.5mm}} % Defines a new command for horizontal lines, change thickness here
	
	\center % Centre everything on the page
	
	%------------------------------------------------
	%	Title
	%------------------------------------------------
	
	% \HRule\\[0.5cm]
	
	% {\huge\bfseries Title Page for }\\[0.4cm] % Title of your document
 %        {\huge\bfseries ``Regression-Based Proximal Causal Inference''}\\[0.4cm] % Title of your document

                \vspace{1cm}
        \begin{center}
            \large
            \begin{itemize}
                \item Title: Regression-based proximal causal inference for right-censored time-to-event data.
                \item Authors: Kendrick Qijun Li, PhD$^{a}$; George C. Linderman, MD, PhD$^b$; Xu Shi, PhD$^{c}$; Eric J. Tchetgen Tchetgen, PhD$^{d}$
                \item Affiliations: \\
                              \begin{tabular}{l}
                $a$: Department of Biostatistics, St. Jude Children’s Research Hospital\\
                $b$: Department of Surgery, Massachusetts General Hospital \\
                $c$: Department of Biostatistics, University of Michigan\\
                $d$: Department of Biostatistics, Perelman School of Medicine, University of Pennsylvania;\\ Department of Statistics and Data Science, Wharton School, University of Pennsylvania 
            \end{tabular}
                \item Correspondence Address: \\
                % \item 
                    – Name: Kendrick Qijun Li \\ 
                    – Mailing address: MS 768, Room S2041, St. Jude Children's Research Hospital, 262 Danny Thomas Place, Memphis, TN 38105-3678 \\
                    – Telephone number: +1 206-816-2786\\
                    – Email address: kendrick.li@stjude.org
                \item Suggested running head: Proximal causal inference for time-to-event data
                
                \item Key words: Negative Control; Survival Analysis; Unmeasured Confounding; Additive Hazards Model
                \item Funding: R01GM139926 (PI: Shi/Tchetgen Tchetgen)\\
                %NIH grants: R01AI127271; R01AG065276; R01GM139926 \\ 
               %PI: Tchetgen Tchetgen, Eric Joel
                \item Conflict of Interest: %N/A
                \item Data Access: The RHC dataset is publicly available at Vanderbilt Biostatistics Datasets webpage (\url{https://hbiostat.org/data/}). Code for simulation and data analysis in this article is available via GitHub (\url{https://github.com/KenLi93/p2sls_surv_manuscript}).
            \end{itemize}
        \end{center}
\noindent

\end{titlepage}

\maketitle
\begin{abstract}
    Unmeasured confounding is a major concern in obtaining credible inferences about causal effects from observational data. \emph{Proximal causal inference} (PCI) is an emerging methodological framework to detect and  potentially account for confounding bias by carefully leveraging a pair of negative control exposure (NCE) and outcome (NCO) variables, also known as treatment and outcome confounding proxies. Although regression-based PCI is well-developed for binary and continuous outcomes, analogous PCI regression methods for right-censored time-to-event outcomes are currently lacking. In this paper, we propose a novel two-stage regression PCI approach for right-censored survival data under an additive hazard structural model. We provide theoretical justification for the proposed approach tailored to different types of NCOs, including continuous, count, and right-censored time-to-event variables. We illustrate the approach with an evaluation of the effectiveness of right heart catheterization among critically ill patients using data from the SUPPORT study. Our method is implemented in the open-access R package `pci2s'.
\end{abstract}

\section{Introduction}
\label{sec:intro}
 
A common task in biomedical or epidemiological research is to study the causal impact of various potential exposures on a time-to-event outcome. Examples include time from the completion of cancer treatment to death or recurrence, time to infection after receiving a vaccine against an infectious disease, time from hospital discharge after major surgery to readmission, etc. Standard statistical methods for right-censored time-to-event outcome data include parametric models (see~\textcite{kleinbaum1996survival}), Cox proportional hazards models~\supercite{cox1972regression}, accelerated failure time models~\supercite{pike1966method,wei1992accelerated}, and additive hazards models~\supercite{aalen1980model,aalen1989linear}, which provide different measures of exposure effects. 

A common challenge for statistical analysis in observational studies is confounding bias induced by common causes of the exposure and outcome variables of interest. To adjust for confounding bias, baseline demographic and clinical factors, such as age, sex, and measures of health status and health related behavior, are usually included in the model. However, important confounders are often not available, particularly in the context of a post-hoc analysis of data collected primarily for health insurance claims or other administrative reasons, and not necessarily for epidemiological or clinical study purposes. For example, an analysis of registry data may fail to include many relevant behavioral and clinical factors.  Furthermore, measured covariates may not fully reflect the complex nature of confounders such as socioeconomic status, patient frailty or health behaviors, resulting in residual confounding~\supercite{kaufman1997socioeconomic,liang2014investigation}. For instance, the most commonly measured indicators of unhealthy behaviors, such as smoking and excessive alcohol consumption, may not fully capture the confounding mechanism of such behaviors when evaluating, say, the causal association between seasonal influenza vaccination and the incidence of flu infection. Such residual confounding bias can be challenging to detect and appropriately correct for.

Throughout the paper, we will consider an application due to \textcite{connors1996effectiveness} who evaluated the association between right heart catheterization (RHC) during the first 24 hours of care in the intensive care unit (ICU) and 30-day mortality among 5,735 critically ill patients in a prospective cohort study. Their analysis found that patients who received RHC during the first 24 hours after study entry had higher 30-day mortality compared with those who didn't (adjusted odds ratio=1.24; 95\% confidence interval: 1.03-1.49). However, their analysis may be subject to unmeasured confounding bias by patients' disease severity: patients with more severe heart conditions were more likely to require RHC and had a higher risk of mortality. As a result, the potential benefits and associated risks of RHC remain unclear without further addressing this potential confounding bias.

Several statistical methods have been proposed  to detect the presence of unmeasured confounding under certain conditions, and to potentially de-bias effect estimates, often under even stronger assumptions. Such approaches include negative control methods~\supercite{lipsitch2010negative}, sensitivity analysis~\supercite{robins2000sensitivity}, instrumental variable methods~\supercite{angrist1996identification}, and difference-in-differences~\supercite{meyer1995natural}. More recently, proximal causal inference (PCI) has emerged as a novel approach to de-bias causal effect estimates~\supercite{miao2018identifying,tchetgen2020introduction}. Proximal causal inference leverages a pair of negative control variables: (i) a negative control outcome (NCO), which is a priori known not to be causally impacted by the treatment in view, but is relevant for the unmeasured confounder in the sense of being associated with the latter; and (ii) a negative control exposure (NCE), which is a priori known not to have a direct causal effect on the outcome of interest, while being relevant for, and therefore associated with the unmeasured confounder.
Factors satisfying either property (i) or (ii) are also sometimes referred to as outcome or treatment confounding proxies, respectively. These variables can be regarded as proxies because conditioning on the unmeasured confounder renders them irrelevant to confounding adjustment. ~(Figure \ref{fig:nc-dag} (a)). 
  Such proxies can therefore sometimes be appropriately viewed as error-prone measurements of the unmeasured confounding mechanism, without necessarily having to specify the nature of the measurement error (e.g., classical measurement error).  This nonparametric view of proxy variables is introduced in \textcite{miao2018identifying} who formally establish conditions under which 
a pair of treatment and outcome confounding proxy variables  nonparametrically identify a causal effect in the presence of unmeasured
confounding.

\textcite{tchetgen2020introduction} generalize PCI to complex longitudinal studies and derive the
proximal g-formula to identify the joint causal effects of time-varying treatments subject to both measured and
latent time-varying confounding, extending James Robins' g-methods to the proximal setting~\supercite{fitzmaurice2008estimation}. They also propose proximal g-computation, a proxy-based generalization of the g-computation algorithm.
In the case of continuous outcomes, under linear models for NCO $W$ and primary outcome $Y$ , \textcite{tchetgen2020introduction} establish that the corresponding proximal g-computation algorithm can be implemented
by following a two-stage regression procedure. Recently, \textcite{liu2024regression} extended the two-stage regression approach to handle binary, polytomous, or count outcomes under 
a familiar generalized linear model formulation in a point exposure setting. As they demonstrate, the regression-based approach is appealing for routine application of PCI as it circumvents the need to solve certain complicated integral equations typically involved in nonparametric PCI estimation~\supercite{ghassami2022minimax}.

Despite these advances, PCI remains somewhat underdeveloped for right-censored failure time outcomes with two notable exceptions; first, \textcite{ying2022proximal} proposed a PCI approach to identify the marginal causal effect of a point exposure on the survival curve; while \textcite{ying2024proximal} subsequently proposed a proximal approach to account for dependent censoring. Similar to the original formulation of \textcite{miao2018identifying} %and \textcite{ying2023proximal}, 
both papers contend with having to solve ill-posed Fredholm integral equations of the first kind to achieve identification and therefore may not be practical in routine epidemiological applications. % However, in practice, researchers are often interested in obtaining a summary measure of an exposure effect. Epidemiological studies often want to study the association between several exposures and outcome at once. Ying's approach also requires complicated statistical programming, which limits its application. 
A simple regression-based PCI approach for survival outcomes akin to the two-stage regression approach of ~\textcite{tchetgen2020introduction} and \textcite{liu2024regression} is still lacking.% and the main objective of the current paper.

In this article, we extend the two-stage regression approach to the survival context, in which one aims to estimate the causal effect of a point treatment subject to unmeasured confounding under a structural additive hazards model for a right-censored time-to-event primary outcome. In Section~\ref{sec:2sls}, we discuss our approach for a single NCO of various type, including continuous, count, binary, right-censored time-to-event outcomes, and possibly a competing risk for the primary endpoint. %NCOs that can be modeled by linear models (continuous variables), log-linear models (continuous, binary, or count variables), additive hazards models (right-censored time-to-event variables).
In Section~\ref{sec:multi-nco}, we further extend the approach to handle multivariable NCOs when available. In Section~\ref{sec:sim}, we demonstrate using simulation studies that the proposed approach can successfully correct for unmeasured confounding bias. In Section~\ref{sec:rhc}, we illustrate our methods using the SUPPORT data of \textcite{connors1996effectiveness} to evaluate the effectiveness of RHC on survival. We defer all proofs and additional discussion to the Appendix.% and the Medicare data to compare the effectiveness of lobar and sublobar resection for stage 1-3 non-small cell lung cancer patients. 

%ETT. YOU FORMALLY STATE THE ASSUMPTIONS IN THE FOLLOWING SECTIONS BUT NOT THE RESULTS. CAN YOU HAVE THE RESULTS STATED FORMALLY. THE STRUCTURE SHOULD BE SIMILAR TO JIEWEN'S PAPER OTHERWISE THE PRESENTATION IS NOT ALWAYS CLEAR AS TO WHAT IS A FORMAL RESULT AND WHAT IS NOT.

\section{Review of the two-stage least-squares approach for continuous outcome and NCO}\label{sec:linear}
We first provide a brief review of the two-stage least-squares approach for continuous outcome and NCO variables in  \textcite{tchetgen2020introduction}. Let $A$ be their exposure of interest, $Y$ be the outcome, and $X$ and $U$ be measured and unmeasured confounders, respectively. Further denote the  negative control exposure and outcome variables as $Z$ and $W$, respectively, which satisfy the following conditional independence conditions %$W\indep A\mid U, X$ and 
$(Z,A)\indep (W,Y^a)\mid  U, X$ for all $a$ in the support of A, formalizing the definition of valid NCE and NCO variables, together with exchangeability or no residual confounding given $(X,U)$. \textcite{tchetgen2020introduction} considered the following linear structural models: 
\begin{align}
	&E(Y|A,Z,U,X) = \beta_0+\beta_AA+\beta_UU + \beta_X^T X \label{c1} \\ 
	&E(W|A,Z,U, X) = c_0 + c_UU + c_X^TX \label{c2}
\end{align}
In the above equations, $Z$ does not appear on the right-hand side of Equation~\eqref{c1}, and $(A,Z)$ do not appear on the right-hand side of Equation~\eqref{c2}, which is compatible with the conditional independence assumptions of the negative control variables. The parameter $\beta_A$  encodes the causal effect of $A$ on $Y$.

Because these models condition on the unobserved $U$,  $\beta_A$ cannot be identified by a standard linear regression model. By evaluating the conditional expectations of $U$ given $(A,Z,X)$ on both sides of both equations, we have:
\begin{align*} 
	&E(Y|A,Z,X) = \beta_0+\beta_AA+\beta_UE(U\mid A,Z,X) + \beta_X^T X\\ 
	&E(W|A,Z,X) = c_0 + c_UE(U\mid A,Z,X) + c_X^TX
\end{align*}
and therefore $$E(Y|A,Z,X) = \beta_0^*+\beta_AA+\beta_U^*E(W\mid A,Z,X) + (\beta_X^*)^T X$$
where $\beta_0^*=\beta_0-\beta_Uc_0/c_U$, $\beta_U^*=\beta_U/c_U$ and $\beta_X^*=\beta_X-\beta_Xc_X/c_U$.  Naturally, this requires $c_U \neq 0$, encoding $U$-relevance of the NCO $W$, that is $W$ must be associated with $U$. The above results indicate that if $E(W\mid A, Z, X)$ were known, one may estimate $\beta_A$ by regressing $Y$ on $(A,E(W|A,Z,X),X)$ via ordinary least squares. In practice, $E(W\mid A,Z,X)$ may be replaced by an estimate obtained via a regression model of $W$ on $(A, Z, X)$. As such, \textcite{tchetgen2020introduction} proposed the following proximal two-stage least-squares (P2SLS) algorithm for estimating $\beta_A$:
\begin{enumerate}
  \item Perform the first-stage regression $W$ on $(A,Z,X)$ using, say, a linear regression model
  $$E(W\mid A,Z,X)=c_0^* + c_A^*A + (c_Z^*)^TZ + (c_X^*)^TX.$$
  \item Compute $\widehat{E}(W\mid A,Z,X)= \widehat c_0^* + \widehat c_A^*A + (\widehat c_Z^*)^TZ + (\widehat c_X^*)^TX$ using the estimated coefficients from the first-stage regression.
  \item Perform the second-stage linear regression for $Y$ with independent variables $A$, $\widehat{E}(W\mid A,Z,X)$ and $X$.
\end{enumerate}

The above algorithm is reminiscent of the familiar two-stage least-squares instrumental variable regression approach and routinely implemented in the standard statistical software, such as the R packages \textit{gmm} and \textit{ivreg}~\supercite{baiocchi2014instrumental}. In an analysis of the RHC data described in Section~\ref{sec:intro}, Tchetgen Tchetgen et al. used the above algorithm to study the effect of RHC on 30-day survival. The outcome was  the number of days between admission and death or censoring at 30 days. The two-stage approach estimated the causal effect to be -1.80 days (standard error: 0.43), while the estimated effect size from the ordinary least square is -1.25 days (standard error: 0.28).  Both results suggested that RHC is associated with higher 30-day mortality~\supercite{tchetgen2020introduction}, although the PCI estimate is more pronounced.

\textcite{liu2024regression} extended the P2SLS algorithm to continuous, binary, and count $W$ and $Y$ via generalized linear models (GLMs) to appropriately account for different variable types.  In the following sections, we shall consider various settings with a time-to-event primary outcome variable and NCOs potentially of different type.

\section{Univariate NCO and unmeasured confounder}
\label{sec:2sls}
In this Section, we introduce our approach for univariate $U$ and NCO, with one or more NCEs.

As before, $U$ and $X$ are  unmeasured and measured baseline confounders for the effect of $A$ on $T$, respectively. We assume that the data are generated under Aalen's semiparametric additive hazards model~\supercite{aalen1980model,aalen1989linear,lin1994semiparametric,mckeague1994partly}:

\begin{assumption}[Additive hazards model]\label{assump:ah0}
    The conditional hazard function $\lambda_T(t\mid A, U, X)$ for the outcome $T$ given $(A, U, X)$, satisfies
    \begin{equation}\label{eq:ah0}
        \lambda_T(t\mid A, U, X)=\beta_0(t) + \beta_A^TA + \beta_X^TX + \beta_UU. 
    \end{equation}
%    where the parameters $\beta_0(\cdot)$, $\beta_A$, $\beta_X$ and $\beta_U$ satisfy \begin{equation}\label{eq:positivity}
%        \beta_0(t) + \beta_A^TA + \beta_X^TX + \beta_U^TU \geq 0
%\end{equation}
   % almost surely for all $t>0$.
\end{assumption}
Assumption~\ref{assump:ah0} states that conditional on the exposure $A$ and confounders $(U, X)$, the conditional hazard function of $T$ at each time $t$ is linear in $(A, X, U)$. In Model~\eqref{eq:ah0}, the exposure of interest $A$ is allowed to be any data type.  %ETT. IM TAKING THIS OUT, WHILE IT IS CORRECT, IT BRINGS UP A WHOLE CAN OF WORMS THAT IS NOT ADDRESSED IN THE PAPER, WHICH IS HOW IS THIS IMPOSED IN THE OBSERVED SAMPLE. I THINK OK TO REMOVE AS WE DO NOT AIM TO ESTIMATE THE BASELINE HAZARD FUNCTION IN OUR APPPROACH AND THE COEFFICIENTS WE DO ESTIMATE ARE UNRESTRICTED. 
Importantly, the baseline hazard function $\beta_0(t)$ is an unrestricted function of time, while the regression coefficients $\beta_A$, $\beta_X$ and $\beta_U$ are a priori unrestricted constants. %other than the natural constraint \eqref{eq:positivity}, which ensures that $\lambda_T(\cdot\mid A, U, X)$ is a proper hazard function.
Due to the relationship between the hazard and survival functions, Model~\eqref{eq:ah0} implies 
\begin{equation}\label{eq:ah-survival}
    \dfrac{P(T>t\mid A, U, X)}{P(T>t\mid A=0, U=0, X=0)}=\exp\{-t(\beta_A^TA + \beta_X^TX + \beta_UU)\}.
\end{equation}

Model~\eqref{eq:ah0} further has a causal interpretation. Suppose $A$ is binary and the standard identifiability assumptions for causal effects hold, including (1) Consistency: $T^{A=a}=T$ if $A=a$ for $a=0, 1$, and (2) Exchangeability: $T^{A=a}\indep A\mid U,X$ for $a=0, 1$, where $T^{A=a}$ is the potential time-to-event outcome if $A=a$, and (3) Positivity: $0< P(A = a\mid U, X) < 1$ almost
surely, for $a = 0, 1$. In Appendix S1.2, we show that Model~\eqref{eq:ah0} is equivalent to the multiplicative counterfactual survival model %for  $S_{T^{A=a}}(t\mid X)$: %for $a=0,1$, the conditional survival functions for the potential time-to-event outcomes given $X$ at time $t$ satisfies
\begin{equation}\label{eq:causal}
    \dfrac{S_{T^{A=1}}(t\mid X)}{S_{T^{A=0}}(t\mid X)} = \exp\left(-\beta_At\right).
\end{equation}
Therefore, we generally refer to the parameter of interest $\beta_A$ as ``causal hazard difference''.

 Assumption~\ref{assump:ah0} states that the conditional hazard function is linear in $(A, U, X)$ and the regression coefficients are constant over time. These conditions can be relaxed  considerably as in Assumption~\ref{assump:ah-star} below: 

\begin{newassumption}{1*}{Additive hazards model}\label{assump:ah-star}
    The conditional hazard function $\lambda_T(t\mid A, U, X)$, satisfies
    \begin{equation}\label{eq:ah-star}
        \lambda_T(t\mid A, U, X)=\beta_0(t,X) + \beta_{A,X}(t, A, X) +  \beta_U(t)U 
    \end{equation}
    where $\beta_{A, X}(t, 0, X)=0$ so that the functions are identified.
   % where  functions $\beta_0(\cdot), \beta_A(\cdot), \beta(\cdot), \beta_X(\cdot)$ satisfy \begin{equation}\label{eq:positivity-star}
    %    \beta_0(t) + \beta_{A,X}(t, A, X) + \beta_U(t)^TU \geq 0
    %\end{equation}
    %almost surely.
\end{newassumption}

In Assumption~\ref{assump:ah-star}, we allow the effect of $(A,X)$ to be time-varying and nonlinear, so that $A$ may have arbitrary interaction with $X$. We also allow the effect of $U$ to be time-varying; however, we impose that $(A, X)$ does not have an additive interaction with the unmeasured confounders $U$ on the hazard scale. Identification and estimation of the causal effect under this more flexible model is further discussed in Appendix S2. The main text will focus primarily on the simpler formulation in Model~\eqref{eq:ah0} to facilitate the exposition.

For each subject, suppose we also observe a negative control outcome (NCO) $W$ that is not directly affected by the treatment, and a negative control exposure $Z$ that does not have a direct effect on $W$, $T$ or $C$. Formally, we make the following assumption:
\begin{assumption}[Negative control variables]\label{assump:nc}
    The NCE $Z$ and NCO $W$ satisfy $A\indep W\mid U, X$ and $Z\indep (W, T)\mid A, U, X$.
\end{assumption}
Assumption~\ref{assump:nc} corresponds to the formal assumptions defining negative control variables (\textcite{miao2018identifying} and \textcite{tchetgen2020introduction}), with the outcome variable now given in terms of the primary time-to-event $T$.  %ETT. THE FIGURE HAS TWO DAGS, IM ASSUMING YOU ARE REFERRING TO THE (B), IF SO PLEASE SPECIFY AND MAYBE MOVE THIS SENTENCE AFTER YOU HAVE DESCRIBED THE ASSUMPTION FOR CENSORING AS WELL 

Potential negative control variables abound in epidemiological studies. \textcite{shi2020selective} provides more examples of possible DAGs compatible with Assumption~\ref{assump:nc}, as well as a comprehensive review of published studies in epidemiology that have made use of negative controls. In the RHC example reported in \textcite{tchetgen2020introduction} and \textcite{cui2023semiparametric}, the authors considered ten biomarker measurements from blood tests within the initial 24 hours in the ICU as potential negative controls, as these measurements may be subject to measurement error and are clearly proxies for underlying disease severity. Among the ten measurements, four variables including $\text{PaO}_2/\text{FiO}_2$, $\text{PaCO}_2$,blood pH and hematocrit which provide critical information on patients' respiratory function, systemic perfusion, and presence of anemia are not only indicative of the patient's prognosis~\supercite{broccard2013making,messina2022partial, mendelson2005biomedical,mondal2024hematocrit}, but are also strongly correlated with both the treatment and the outcome. Therefore, following the same rationale, we selected NCE and NCO from these four variables.

%ETT. CAN YOU GIVE EXAMPLES OF NCs AT THIS POINT. ALSO IT WOULD BE GOOD TO PROVIDE A DAG FOR PCI, IN FACT EITHER HERE OR INTRO YOU CAN REFER BACK TO. 

\begin{figure}[!htbp]
		\centering
        \begin{tabular}{cc}
        \resizebox{.47\textwidth}{!}{
            \begin{tikzpicture}
			
			\tikzset{line width=1pt,inner sep=5pt,
				%	%swig vsplit={gap=3pt, inner line width right=0.4pt},
				ell/.style={draw, inner sep=1.5pt,line width=1pt}}

			\node[shape = circle, ell] (A) at (-2, 0) {$A$};
			\node (Aname) at (-2.4, 0.5-1) {Exposure};

			\node[shape=ellipse,ell] (Y) at (2,0) {$Y$};
			\node (Yname) at (1.6,0.5-1) {Outcome};

			\node[shape=ellipse,ell] (U) at (0,1.5+1) {$U$};
			%			\node (HSname) at (0,1.5+0.5+0.5) {Unmeasured};
			\node (HSname) at (0,1.5+1.7) {Unmeasured confounders};

			%%%NCs
			\node[shape=ellipse,ell] (Z) at (-2.7,1.5) {$Z$};
                \node (Zname1) at (-2.9, 2.5) {Negative control exposure};
			\node (Zname) at (-2.9, 2.1) {(NCE)};
			
			\node[shape=ellipse,ell] (W) at (2.7,1.5) {$W$};
                \node (Wname1) at (2.9, 2.5) {Negative control outcome};
			\node (Wname) at (2.9, 2.1) {(NCO)};
			\draw[dashed,stealth-stealth, line width=0.5pt](Z) to (A);
			
			\draw[-stealth, line width = 0.5pt] (W) to (Y); %% the arrow can actually be both direction!!

			\draw[-stealth,line width=0.5pt](A) to (Y);
			
			\foreach \from/\to in {U/A, U/Y, U/Z, U/W}
			\draw[-stealth, line width = 0.5pt] (\from) -- (\to);
			
		\end{tikzpicture}}&
		\resizebox{.47\textwidth}{!}{\begin{tikzpicture}
			
			\tikzset{line width=1pt,inner sep=5pt,
				%	%swig vsplit={gap=3pt, inner line width right=0.4pt},
				ell/.style={draw, inner sep=1.5pt,line width=1pt}}

			\node[shape = circle, ell] (A) at (-2.5, 0) {$A$};
			\node (Aname) at (-3.7, 0) {Exposure};

			\node[shape=ellipse,ell] (Y) at (0,-1) {$T$};
			\node (Yname) at (1.6,-1) {Time to event};

			\node[shape=ellipse,ell] (U) at (0,1.5+1) {$U$};
			%			\node (HSname) at (0,1.5+0.5+0.5) {Unmeasured};
			\node (HSname) at (0,1.5+1.7) {Unmeasured confounders};			
			
			%%%NCs
			\node[shape=ellipse,ell] (Z) at (-2.7,1.5) {$Z$};
                \node (Zname1) at (-2.9, 2.5) {Negative control exposure};
			\node (Zname) at (-2.9, 2.1) {(NCE)};
			
			\node[shape=ellipse,ell] (W) at (2.7,1.5) {$W$};
                \node (Wname1) at (2.9, 2.5) {Negative control outcome};

                \node[shape=ellipse,ell] (C) at (2.5, 0) {$C$};
			\node (Cname) at (3.7, 0) {Censoring};

			\node (Wname) at (2.9, 2.1) {(NCO)};
			\draw[dashed,stealth-stealth, line width=0.5pt](Z) to (A);
			
			\draw[-stealth, line width = 0.5pt] (W) to (Y); %% the arrow can actually be both direction!!

			\draw[-stealth,line width=0.5pt](A) to (Y);

			\foreach \from/\to in {U/A, U/Y, U/Z, U/W, Z/C, A/C}
			\draw[-stealth, line width = 0.5pt] (\from) -- (\to);
			
		\end{tikzpicture}}\\
        (a) & (b)
        \end{tabular}
	\caption{\label{fig:nc-dag}Possible directed acyclic graphs (DAGs) of the causal relationship for variables that satisfy the required negative control independence assumptions. Dashed double-headed arrows indicate effects that may operate on either direction. Conditioning of measured confounders $X$ is implicit.}
		%			%vaccination status (A),  infection status (Y), testing (T),  latent healthcare-seeking behavior (H), and  unmeasured confounders (X).
		%		}
	\end{figure}

  A well-known challenge with time-to-event outcome is the presence of right censoring, which we now address. For each subject, let $T^*=\min(T,C)$ denote the observed censored event time with $C$ the underlying censoring time. We let $\Delta=\mathbbm 1(T\leq C)$ denote the primary event indicator, such that the observed data constitutes independent and identically distributed (iid) samples of $(A, Z, W, X, T^*, \Delta)$. We next state a standard conditional independent censoring assumption that facilitates the identification and estimation of various conditional hazard functions of the primary outcome despite it being subject to censoring:% the parameters in the additive hazard regression models of Equations~\eqref{eq:second-stage-lm-final}, \eqref{eq:second-stage-glm-final} and \eqref{eq:second-stage-ah-final} below can be identified in the presence of right censoring.
\begin{assumption}[Conditionally independent censoring]\label{assump:censoring}$C\indep T\mid A, X, Z$.
\end{assumption}Assumption~\ref{assump:censoring} is a standard assumption in survival analysis literature and states that given the observed treatment and other covariates, the potential censoring time $C$ and time-to-event $T$ are conditionally independent.~\supercite{laan2003unified, rotnitzky2005inverse}  %ETT I BELIEVE THE VAN IN VAN DER LAAN IS LOWER CASE .
Figure~\ref{fig:nc-dag}(b) shows a possible directed acyclic graph (DAG) with negative control variables and censoring mechanism satisfying Assumptions~\ref{assump:nc} and \ref{assump:censoring}. Unlike other variables on the DAG, censoring is assumed not to be directly associated with either the unmeasured confounder $U$ nor the failure time outcome $T$ other than possibly through $(A,X,Z)$, such that conditioning on the latter must render $T$ and $C$ independent. This is a relatively standard condition in the analysis of censored time-to-event outcome, however, the condition may not be reasonable if loss of follow-up is induced by an unmeasured confounder beyond observed covariates. For example, in the RHC study, Assumption~\ref{assump:censoring} may not hold if  patients with more severe illness were more likely to drop out, and such an association is not completely accounted for by conditioning on measured covariates.

Under the stated conditions, we have that 
\begin{equation}\label{eq:ah-survival2}
    \dfrac{P(T>t\mid A, U, X, Z)}{P(T>t\mid A=0, U=0, X=0, Z)}=\exp\{-t(\beta_A^TA + \beta_X^TX + \beta_UU)\}
\end{equation}
with $P(T>t\mid A=0, U=0, X=0) = \exp\{-B_0(t)\}$ and $B_0(t)=\int_0^t \beta_0(u)du$.

  The parameters in Equations~\eqref{eq:ah0} and \eqref{eq:ah-survival2} are not identified due to the unmeasured confounder $U$. To proceed, we make an additional assumption regarding the conditional distribution of the unmeasured confounder:

\begin{assumption}(Location-shift model for $U$)\label{assump:u-model0}
    \begin{equation}\label{eq:u-model0}U=E(U\mid A, Z, X)+\epsilon\end{equation} where the distribution of $\epsilon$ is unrestricted other than $E(\epsilon)=0$ and $\epsilon\indep (A, Z, X)$.
\end{assumption}
%ETT. In above assumption, state that the distribution of epsilon is unrestricted.

Assumption~\ref{assump:u-model0} states that the conditional distribution of $U$ given $(A, Z, X)$ follows a location-shift model, and therefore depends on the latter only through its mean, so that the residual error $\epsilon$ is independent of $(A, Z, X)$. The use of the location-shift model for latent factors has gained prominence in causal inference as several standard distributions, e.g. Gaussian homoscedastic errors satisfy this restriction~\supercite{tchetgen2015instrumental,liu2024regression}. In the RHC example of Section~\ref{sec:rhc}, if disease severity is the unmeasured confounder of concern, Assumption~\ref{assump:u-model0} would require that the variation of the underlying disease severity is homoscedastic, i.e. approximately the same across strata defined by $A, Z, X$.

We establish the following result, proved in Appendix S1:

\begin{result}\label{result:second-stage}
    Under Assumption \ref{assump:u-model0}, Equation~\eqref{eq:ah-survival2} implies the following multiplicative model for the \emph{observed} survival function of $T$ conditional on $(A,Z,X)$:
    \begin{equation}\label{eq:second-stage}
        P(T>t\mid A, Z, X)=\exp\{-\widetilde B_0(t) - t\beta_A^TA - t\beta_UE(U\mid A, Z, X)-t\beta_X^T X\}
    \end{equation}
    where $\widetilde B_0(t) = \int_0^t\beta_0(r)dr -\log\int \exp(-t\beta_Ue)\,\mathrm dF_\epsilon(e)$ and $F_\epsilon$ is the distribution function of $\epsilon$. Furthermore, this is equivalent to an additive hazards model
\begin{equation}\label{eq:second-stage-ah}
    \lambda_T(t\mid A, Z, X)=\widetilde \beta_0(t) + \beta_A^TA + \beta_UE(U\mid A, Z, X) + \beta_X^TX. 
\end{equation}
where $\widetilde \beta_0(t)=\beta_0(t) - \partial\left\{ \log\int \exp(-t\beta_Ue)\,\mathrm dF_\epsilon(e)\right\} / \partial t$. 
\end{result}

By Result~\ref{result:second-stage}, if we can somehow estimate $E(U\mid A, Z, X)$, then the parameter of interest $\beta_A$ can be estimated by additive hazards regression with predictors $A$, $X$ and $E(U\mid A, Z, X)$.

Under a standard model specification, we further suppose that  $E(U\mid A, Z, X)$ follows a linear model:
\begin{assumption}[Linear mean model for $U$]\label{assump:u-lm0}
\begin{equation}\label{eq:u-lm0}
E(U\mid A, Z, X)=\gamma_{0}+\gamma_{A}^TA + \gamma_{Z}^TZ + \gamma_{X}^TX.
\end{equation}
\end{assumption}
We defer to Appendix S2, the more general case where $E(U\mid A, Z, X)=\gamma(A, Z, X)$ is modeled more flexibly allowing for  non-linearity and interactions involving $A$, $Z$ or $X$, by incorporating say  spline basis functions of the variables~\supercite{de1972calculating}.

%ETT. SIMILAR TO JIEWEN'S PAPER, I THINK YOU SHOULD A BRIEF SECTION OR SUBSECTION TO REVIEW THE LINEAR MODEL for Y CASE BEFORE THE SURVIVAL CASE.
Inspired by the two-stage-least-squares approach in \textcite{tchetgen2020introduction}, our approach for estimating $E(U\mid A, Z, X)$ leverages the NCO $W$ -- we obtain an estimator of the latter using either  a linear, multiplicative or additive hazards regression model for $W$ conditional on $(A,Z,X)$, depending on $W$'s data type. In each case, we establish that under our model specification, the fitted value for $W$ recovers $E(U\mid A, Z, X)$ up to a linear transformation and therefore the former can be used as a substitute for the latter in the additive hazards regression model for the primary outcome together with $(A, X)$ to produce consistent estimators of $\beta_A$ and $\beta_X$.

%This is a fairly standard assumption in survival analysis ETT. CITE STANDARD SURVIVAL TEXTBOOKS, YING JUST USES THIS AS WELL I DONT THINK YOU NEED TO CITE HIM FOR THIS AS PRECEEDS HIM IN THE LITERATURE 
%standathe same requirement for the censoring mechanism in \textcite{ying2024proximal}.

%Below we discuss the identification and estimation of $E(U\mid A, Z, X)$ up to a multiplicative constant for different types of $W$.

\subsection{$W$ follows a linear model}\label{sec:w-linear}
We first consider the simple scenario where $W$ follows a linear mean model: 

\begin{newassumption}{6A}{NCO follows a linear mean model}\label{assump:w-linear}
\begin{equation}\label{eq:w-linear}
    E(W\mid A, U, Z, X)=c_{01} + c_{U1}U+c_{X1}^TX.
\end{equation}
\end{newassumption}

Note that while in Equation~\eqref{eq:w-linear}, both $A$ and $Z$ appear in the conditional event on the left-hand side, consistent with Assumption~\ref{assump:nc}, they do not appear on the right-hand side of the equation. %ETT. IM REMOVING YOUR EXAMPLE AS IT IS LITERALLY THE DEFINITION OF EQUATION (10) NOT AN EXAMPLE OF IT% For example, Assumption~\ref{assump:w-linear} holds if $W=c_0 + c_U^TU+c_X^TX+\epsilon_W$ where the residual error $\epsilon_W$ satisfies $E(\epsilon_W\mid A, U, Z, X)=0$. 

Equation~\eqref{eq:w-linear} immediately implies that
\begin{equation}\label{eq:first-stage-linear}
E(W\mid A, Z, X)=c_{01} + c_{U1}E(U\mid A, Z, X) + c_{X1}^TX.
\end{equation}

Therefore, we have the following result:

\begin{newresult}{2A}{}\label{result:id-linear}
Under Assumptions~\ref{assump:ah0}-\ref{assump:u-lm0}, \ref{assump:w-linear} and \ref{assump:completeness-lin} below, $\beta_A$ can be identified with the following equations:
\begin{align}
    E(W\mid A, Z, X)&=c_{01}^* + (c_{A1}^*)^T A + (c_{Z1}^*)^TZ + (c_{X1}^*)^TX\label{eq:first-stage-lm-final}\\
    \lambda_T(t\mid A, Z, X) &=  \beta_{01}^{*}(t)+\beta_A^TA + \beta_{U1}^*E(W\mid A, Z, X) + (\beta_{X1}^*)^TX\label{eq:second-stage-lm-final}
\end{align}
where  $c_{01}^*=c_{01} +c_{U1}\gamma_{0}$, $c_{A1}^*=c_{U1}\gamma_{A}$, $c_{Z1}^*=c_{U1}\gamma_Z$, $c_{X1}^*=c_{U1}\gamma_X + c_{X1}$, $\beta_{01}^*(t) = \widetilde\beta_0(t) - \beta_Uc_{01}/c_{U1}$, $\beta_{U1}^* = \beta_U/c_{U1}$, and $\beta_{X1}^*=\beta_X - \beta_Uc_{X1}/c_{U1}$.  
\end{newresult}

%ETT. I DONT THINK IT IS A GOOD IDEA TO FORMULATE ASSUMPTION 7 IN TERMS OF COMPLETENESS AS IT IS ACTUALLY NOT NEEDED UNDER THE PARAMETRIC MODEL FORMULATION WHICH IS A REAL ADVANTAGE OF THE APPROACH. PLEASE REMOVE THIS. ALL YOU NEED AS FAR AS I CAN TELL IS THAT Z AND W ARE RELEVANT FOR U SUCH THAT E(W|A,Z,X) DEPENDS ON Z. THIS ALLOWS US TO SEPARATE THE CAUSAL EFFECT OF A FROM THE SPURIOUS ASSOCIATION W A VIA U (UPON COLLAPSING OF W) BECAUSE OF THE VARIABILITY IN Z WHICH ONLY APPEARS VIA THE MODEL FOR W. THIS DESERVES FAR MORE CAREFUL DISCUSSION.

From the above exposition, we see that the coefficients in the regression models~\eqref{eq:first-stage-lm-final} and~\eqref{eq:second-stage-lm-final} correspond to the parameters in the underlying model of Equation~\eqref{eq:first-stage-linear} only if $c_{U1}\neq 0$, formalizing the requirement that $W$ needs to be relevant for $U$. On the other hand, if $\gamma_Z=0$, then $E(W\mid A, Z, X)$ does not depend on $Z$ and therefore may become perfectly co-linear with $A$ and $X$ such that $\beta_A$ cannot be identified from Equation~\eqref{eq:second-stage-lm-final}. As such, another required condition for identification of $\beta_A$ is $\gamma_Z\neq 0$, which indicates that $Z$ also needs to be relevant to $U$. This allows us to separate the causal effect of $A$ from the unmeasured confounding bias via $U$ through the variability in $Z$, which only appears via the model for $W$. 

We state the above two conditions as below:

\begin{newassumption}{7A}{U-relevance of negative controls}\label{assump:completeness-lin}
   The parameters in Equations~\eqref{eq:u-lm0} and \eqref{eq:w-linear} satisfy $c_{U1}\neq 0$ and $\gamma_Z\neq 0$.
\end{newassumption}
 %ETT. THIS WILL NEED TO BE RESTATED ONCE THE ASSUMPTION IS STATED IN TERMS OF RELEVANCE.I THINK ALSO THIS SHOULD NOT BE CONDITIONING ON $Y=0,S=1$ RIGHT, MAYBE FROM CUT AND PASTE FROM THE TND PAPER?

As all random variables in Equations~\eqref{eq:first-stage-lm-final} and \eqref{eq:second-stage-lm-final} are observed, $\beta_A$ can be identified. Result~\ref{result:id-linear} motivates the following two-stage regression method, which we refer to as Proximal Two-Stage Regression for Survival data (P2SR-Surv) and summarized in Algorithm~\ref{alg:p2sls-surv-linear}.
\begin{algorithm}[!h]
    \caption{Proximal Two-Stage Regression for Survival data (P2SR-Surv) with a linear NCO}
    \label{alg:p2sls-surv-linear}
    \begin{algorithmic}[1]
        \State Fit the linear regression model according to Equation~\eqref{eq:first-stage-lm-final} and obtain the estimators for the regression coefficients $\widehat c_{01}^*$, $\widehat c_{A1}^*$, $\widehat c_{Z1}^*$ and $\widehat c_{X1}^*$;
        \State Obtain the linear predictors $$\widehat\mu_1(A, Z, X)=\widehat c_{01}^* + (\widehat c_{A1}^*)^TA + (\widehat c_{Z1}^*)^TZ + (\widehat c_{X1}^*)^TX.$$
        \State Fit the additive hazards regression model according to Equation~\eqref{eq:second-stage-lm-final} with $E(W\mid A, Z, X)$ replaced by $\widehat\mu_1(A, Z, X)$. The regression coefficient for $A$ is an estimator of $\beta_A$.
    \end{algorithmic}
\end{algorithm}

Inference for $\beta_A$ appropriately accounting for the uncertainty at both stages of estimation may be based on generalized method of moments (See Appendix S3) or nonparametric bootstrap~\supercite{efron1994introduction}.

\subsection{$W$ follows a GLM with log link function}
%ETT. I THINK THIS SECTION SHOULD ALSO BE ABOUT UNIVARIATE W AND THEREFORE U.
Generalized linear models with log link function posit that the conditional mean for an outcome conditional on covariates is linear on the exponential scale. Examples of such log-linear regression models include Poisson or Negative Binomial regression models, Gamma regression models, and log-binomial models, commonly used to model count outcomes, and more generally, outcomes a priori known to be nonnegative%continuous variables, or binary variables, respectively 
~\supercite{nelder1972generalized}. Suppose the distribution of $W$  follows such a GLM with log link: 
\begin{newassumption}{6B}{NCO follows a log-linear mean model}\label{assump:w-loglinear}
\begin{equation}\label{eq:w-loglinear}
    E(W\mid A, U, Z, X)=\exp\{c_{02} + c_{U2}U +c_{X2}^T X\}
\end{equation}

\end{newassumption}

Similar to Assumption~\ref{assump:w-linear}, $A$ and $Z$ do not appear on the right-hand side of Equation~\eqref{eq:w-loglinear}, consistent with Assumption~\ref{assump:nc}. An offset may be included in the model to account for differential follow-up time during which $W$ is measured across units. For example, if $W$ is cumulative count of events experienced during a time period, the logarithm of the duration will typically be included as an offset~\supercite{nelder1972generalized}.

As before, we assume the negative control variables are $U$-relevant. More formally:

\begin{newassumption}{7B}{U-relevance of negative controls}\label{assump:completeness-loglin}
   The parameters in Equations~\eqref{eq:u-lm0} and \eqref{eq:w-loglinear} satisfy $c_{U2}\neq 0$ and $\gamma_Z\neq 0$.
\end{newassumption}

We obtain the following result:

\begin{newresult}{2B}{}\label{result:id-loglinear}
Under Assumptions~\ref{assump:ah0}-\ref{assump:u-lm0}, \ref{assump:w-loglinear} and \ref{assump:completeness-loglin}, $\beta_A$ can be identified by through the following equations:
\begin{align}
    E(W\mid A, Z, X) &= \exp\{ c_{02}^{*} + (c_{A2}^*)^T A + (c_{Z2}^*)^T Z + (c_{X2}^*)^TX\}\label{eq:first-stage-glm-final}\\
    \lambda_T(t\mid A, Z, X) &= \beta_{02}^{*}(t)+\beta_A^TA + \beta_{U2}^*\mu_2(A, Z, X) + (\beta_{X2}^*)^TX\label{eq:second-stage-glm-final}
\end{align}
where  $c_{02}^*=c_{02} +\log\left\{ \int\exp(c_{U2}e)dF_\epsilon(e)\right\} + c_{U2}\gamma_{0}$, $c_{A2}^*=c_{U2}\gamma_{A}$, $c_{Z2}^*=c_{U2}\gamma_Z$, $c_{X2}^*=c_{U2}\gamma_X + c_{X2}$, $\beta_{02}^*(t) = \widetilde\beta_0(t) - \beta_U\left\{\widetilde c_{02}+\log\left\{ \int\exp(c_{U2}e)dF_\epsilon(e)\right\}\right\}/c_{U2}$, $\beta_{U2}^* = \beta_U/c_{U2}$, $\beta_{X2}^*=\beta_X - \beta_Uc_{X2}/c_{U2}$, and $$\mu_2(A, Z, X) = c_{02}^{*} + (c_{A2}^*)^T A + (c_{Z2}^*)^T Z + (c_{X2}^*)^TX.$$
\end{newresult}

Result~\ref{result:id-loglinear} also suggests a two-stage regression method for estimating $\beta_A$ (Algorithm~\ref{alg:p2sls-surv-loglin}).

\begin{algorithm}[!h]
    \caption{P2SR-Surv with a log-linear NCO}
    \label{alg:p2sls-surv-loglin}
    \begin{algorithmic}[1]
        \State Fit a log-linear regression model according to Equation~\eqref{eq:first-stage-glm-final} and obtain the estimators for the regression coefficients $\widehat c_{02}^*$, $\widehat c_{A2}^*$, $\widehat c_{Z2}^*$ and $\widehat c_{X2}^*$;
        \State Obtain the linear predictors $$\widehat\mu_2(A, Z, X)=\widehat c_{02}^* + (\widehat c_{A2}^*)^TA + (\widehat c_{Z2}^*)^TZ + (\widehat c_{X2}^*)^TX.$$
        \State Fit the additive hazards regression model according to Equation~\eqref{eq:second-stage-glm-final} with $\mu_2(A, Z, X)$ replaced by $\widehat\mu_2(A, Z, X)$. The regression coefficient for $A$ is an estimator of $\beta_A$.
    \end{algorithmic}
\end{algorithm}

\subsection{$W$ follows an additive hazards model} 
Finally, we consider the case where $W$ is a time-to-event variable that is subject to the same censoring mechanism as the primary time-to-event outcome. We assume that the distribution of $W$ follows an additive hazards model:

\begin{newassumption}{6C}{NCO follows a linear additive hazards model}\label{assump:w-ah}
\begin{equation}\label{eq:w-ah}
    \lambda_W(t\mid A, U, Z, X)=c_{03}(t) + c_{U3}U+c_{X3}^TX.
\end{equation}
Moreover, in addition to Assumption~\ref{assump:censoring}, the censoring mechanism also satisfies $C\indep W\mid A, Z, X$.
\end{newassumption}

We assume $W$ and $T$ are censored concurrently.
In other words, the collection of a subject's information on $W$ and $T$ is terminated at the same time. This may occur, for example, if the censoring is due to subject dropout, so that all information after the dropout time is unavailable.  Similar to Assumption~\ref{assump:censoring}, we assume that the censoring and $W$ are conditionally independent given the other measured variables $(A, Z, X)$. Assumptions~\ref{assump:censoring} and \ref{assump:w-ah} jointly implies $C\indep (W, T)\mid A, Z, X$. Suppose the U-relevance of negative control variables also holds:

\begin{newassumption}{7C}{U-relevance of negative controls}\label{assump:completeness-ah}
   The parameters in Equations~\eqref{eq:u-lm0} and \eqref{eq:w-ah} satisfy $c_{U3}\neq 0$ and $\gamma_Z\neq 0$.
\end{newassumption}

Identification of $\beta_A$ is implied via the following result:

\begin{newresult}{2C}{}\label{result:id-ah}
Under Assumptions~\ref{assump:ah0}-\ref{assump:u-lm0}, \ref{assump:w-ah} and \ref{assump:completeness-ah}, $\beta_A$ can be identified by through the following equations:
\begin{align}
    \lambda_W(t\mid A, Z, X) &= c_{03}^*(t) + (c_{A3}^*)^TA + (c_{Z3}^*)^TZ + (c_X^*)^TX\label{eq:first-stage-ah-final}\\
    \lambda_T(t\mid A, Z, X) &= \beta_{03}^{*}(t) + \beta_A^TA + (\beta_{U3}^*)^T\mu_3(A, Z, X) + (\beta_{X3}^*)^TX\label{eq:second-stage-ah-final}
\end{align}
where  $c_{03}^*=c_{03}(t)-\partial\left\{ \log\int \exp(-tc_{U3}e)dF_\epsilon(e)\right\}/\partial t + c_{U3}\gamma_0$, $c_{A3}^*=c_{U3}\gamma_{A}$, $c_{Z3}^*=c_{U3}\gamma_Z$, $c_{X3}^*=c_{U3}\gamma_X + c_{X3}$, $\beta_{03}^*(t) = \widetilde\beta_0(t)+\beta_U\gamma_0$, $\beta_{U3}^* = \beta_U/c_{U3}$, $\beta_{X3}^*=\beta_X - \beta_Uc_{X3}/c_{U3}$, and $$\mu_3(A, Z, X) = (c_{A3}^*)^T A + (c_{Z3}^*)^T Z + (c_{X3}^*)^TX.$$
\end{newresult}

By Result~\ref{result:id-ah}, a two-stage regression method can be used to estimate $\beta_A$ (Algorithm \ref{alg:p2sls-surv-ah}).

\begin{algorithm}[!htbp]
    \caption{P2SR-Surv with a time-to-event NCO modeled by an additive hazards model}
    \label{alg:p2sls-surv-ah}
    \begin{algorithmic}[1]
        \State Fit an additive hazards regression model according to Equation~\eqref{eq:first-stage-ah-final}  and obtain the estimators for the regression coefficients $\widehat c_{A3}^*$, $\widehat c_{Z3}^*$ and $\widehat c_{X3}^*$;
        \State  Obtain the linear predictors $$\widehat\mu_3(A, Z, X)=(\widehat c_{A3}^*)^TA + (\widehat c_{Z3}^*)^TZ + (\widehat c_{X3}^*)^TX.$$
        \State Fit the additive hazards regression model according to Equation~\eqref{eq:second-stage-ah-final} with $\mu_3(A, Z, X)$ replaced by $\widehat\mu_3(A, Z, X)$. The regression coefficient for $A$ is an estimator of $\beta_A$.
    \end{algorithmic}
\end{algorithm}

 A similar two-stage regression approach can apply to the setting where the event of interest and the negative control event are competing risks. For example, when we compare the cancer-specific mortality between lung cancer patients who receive lobectomy or sublobar resection, death due to stroke may act as a negative control that is a competing risk, assuming that the difference in stroke-related mortality between the two lung surgery procedures is negligible.

 We leave further discussion to Appendices S4 and S5.

\section{Multiple negative control outcomes}\label{sec:multi-nco}
Up to now, we have focused on an univariate unmeasured confounder setting. In practice, the source of unmeasured confounding may be multifaceted. For example, sources of confounding bias for the effect of an exposure on health outcomes can include socioeconomic status, underlying health, access to healthcare, lifestyle, etc.~\supercite{greenland2001confounding}. In other words, $U$ may be multidimensional. This motivates the use of multiple negative control outcomes to capture the complex source of unmeasured confounding. 

To ground ideas, suppose $W=(W^1,\dots, W^{n_W})^T$ is a vector of $n_W$ negative control outcome variables. Extension from the previous sections is straightforward and can be summarized as Algorithm~\ref{alg:p2sls-surv-multi-nco}.

\begin{algorithm}[!htbp]
    \caption{P2SR-Surv with multiple NCOs}
    \label{alg:p2sls-surv-multi-nco}
    \begin{algorithmic}[1]
        \State For each $j=1,\dots, n_W$, fit an appropriate regression model for $W^j$ with independent variables $(A, Z, X)$ according to Equation~\eqref{eq:first-stage-lm-final}, \eqref{eq:first-stage-glm-final} or \eqref{eq:first-stage-ah-final}, depending on the data type of $W^j$;
        \State Obtain the linear predictors in the above regression models $\mu^j(A, Z, X)$, $j=1,\dots, n_W$.
        \State Fit an additive hazard regression model
    $$\lambda_T(t\mid A, Z, X)=\beta_{m0}^*(t) + \beta_A^TA + (\beta_{m1}^*)^T\mu^1(A, Z,X)+\dots +(\beta_{m,n_W}^*)^T\mu^{n_W}(A, Z,X) + (\beta_{mX}^*)^TX.$$
    \end{algorithmic}
\end{algorithm}

We leave a detailed discussion to Appendix S6.

\section{Counterfactual marginal survival function}\label{sec:survfunc}
With the estimate of the causal hazard difference $\beta_A$ in Model~\eqref{eq:ah0} under our assumptions,% if the additional assumptions of consistency and exchangeability hold, 
it is also possible to obtain an estimate of the counterfactual marginal survival function $S_a(t):=P(T^{A=a}>t)$ for each value of $a$ over $t$, as demonstrated in Result~\ref{result:survfunc} below.
%ETT. THESE ASSUMPTIONS ARE EXACTLY THE SAME AS ABOVE, IF SO WHY NOT REFER TO THOSE AND NOT RESTATE THEM
\begin{newresult}{3}{}\label{result:survfunc}Under Assumptions~\ref{assump:ah0} and \ref{assump:nc}, and the additional assumptions of consistency, exchangeability and positivity described in Section~\ref{sec:2sls}, the counterfactual marginal survival function is
    \begin{equation}\label{eq:survfunc}
        S_a(t) =E\{\exp(-t\beta_A^Ta + t\beta_A^T A)P(T>t\mid A, X, Z)\}
    \end{equation}
\end{newresult}

By Sections \ref{sec:2sls}-\ref{sec:multi-nco}, an estimator for $\beta_A$, denoted as $\widehat \beta_A$, can be obtained through Algorithms 1-4, depending on the types of the NCO. Under Assumptions \ref{assump:ah0}, \ref{assump:u-model0} and \ref{assump:u-lm0}, an estimator of $P(T>t\mid A, X,Z)$, denoted as $\widehat P(T>t\mid A, X,Z)$, can be obtained from the corresponding additive hazard estimated model. Therefore, an estimator of $S_a(t)$ is
$$\widehat S_a(t) = \dfrac{1}{n}\sum_{i=1}^n\exp(-t\widehat\beta_A^Ta + t\widehat\beta_A^TA_i)\widehat P(T>t\mid A_i, X_i, Z_i).$$

A counterfactual marginal survival function obtained as above may not be a valid survival function, as it is not restricted to be non-increasing in $t$ nor restricted between 0 and 1, especially in small samples. Post hoc fixes may be applied. For example, following \textcite{lin1994semiparametric}, we may estimate $S_a(t)$ with \begin{equation}\label{eq:survcorrect}\widetilde S_a(t) = \min\{\min_{s\leq t} S_a(t), 1\},\end{equation}
which satisfies the required conditions for a survival function. Pointwise confidence bands of $\widetilde S_a(t)$ may be obtained by nonparametric bootstrap~\supercite{efron1994introduction}.
\section{Simulation study}\label{sec:sim}

We perform simulation studies to evaluate the performance of the proposed methods for unmeasured confounding adjustment. We generate independent and identically distributed data for 1,000 subjects according to an exponential additive hazards regression model, i.e.
$T\mid A, U, X\sim \text{Exponential}(0.2 + 0.2 A + \beta_U U + 0.2 X).$
Here the exposure effect is $\beta_A=0.2$. We vary the unmeasured confounding effect $\beta_U$ between 0 and 2 for different magnitudes of unmeasured confounding bias. 
For simplicity, we consider univariate confounders $U$ and $X$ as independent uniformly distributed variables between 0 and 1. We consider a binary exposure $A$ that follows a logistic regression model given $(U,X)$:
$A\mid U,X \sim \text{Bernoulli}\left(1 / \{1 + \exp(-3 + 5U + X)\}\right).$
Finally, we generate the NCE $Z$ and NCO $W$ both as bivariate Gaussian random variables that follow a linear regression model given $(U,X)$:
\begin{align*}
W\mid U, X\sim N\left(\begin{pmatrix}
            0.5c_U U + 0.2X\\
            2c_U U + X
        \end{pmatrix},\begin{pmatrix}
            0.1^2 & 0 \\
            0 & 0.25^2
        \end{pmatrix}\right), \, Z\mid U, X\sim N\left(\begin{pmatrix}
            c_U U + 0.5X\\
            0.5c_U U + 2X
        \end{pmatrix},\begin{pmatrix}
            0.5^2 & 0 \\
            0 & 0.2^2
        \end{pmatrix}\right)
\end{align*}
where $c_U$ indicates the association between the proxies and unmeasured confounders and is set to be $1$, $0.2$ or $0$. The simulation settings are summarized in Appendix S7. We compare three methods: (1) the proposed proximal two-stage regression method (P2SR-Surv), where each entry in $W$ is modeled with a linear regression model, and the linear predictors are included in an additive hazards regression model for $T$, according to Sections~\ref{sec:w-linear} and \ref{sec:multi-nco}; (2) the ``na\"ive'' additive hazards regression model with only $A$ and $X$ as independent variables, and (3) the ``fully-adjusted'' additive hazards regression model where the independent variables include $A$, $X$, $W$ and $Z$. We used the partial-likelihood-based approach in \textcite{lin1994semiparametric} and \textcite{mckeague1994partly} for estimation and inference of the additive hazards regression models.

\begin{figure}[!htbp]
        \centering      
        \begin{tabular}{cc}
         \includegraphics[width = 0.45\textwidth]{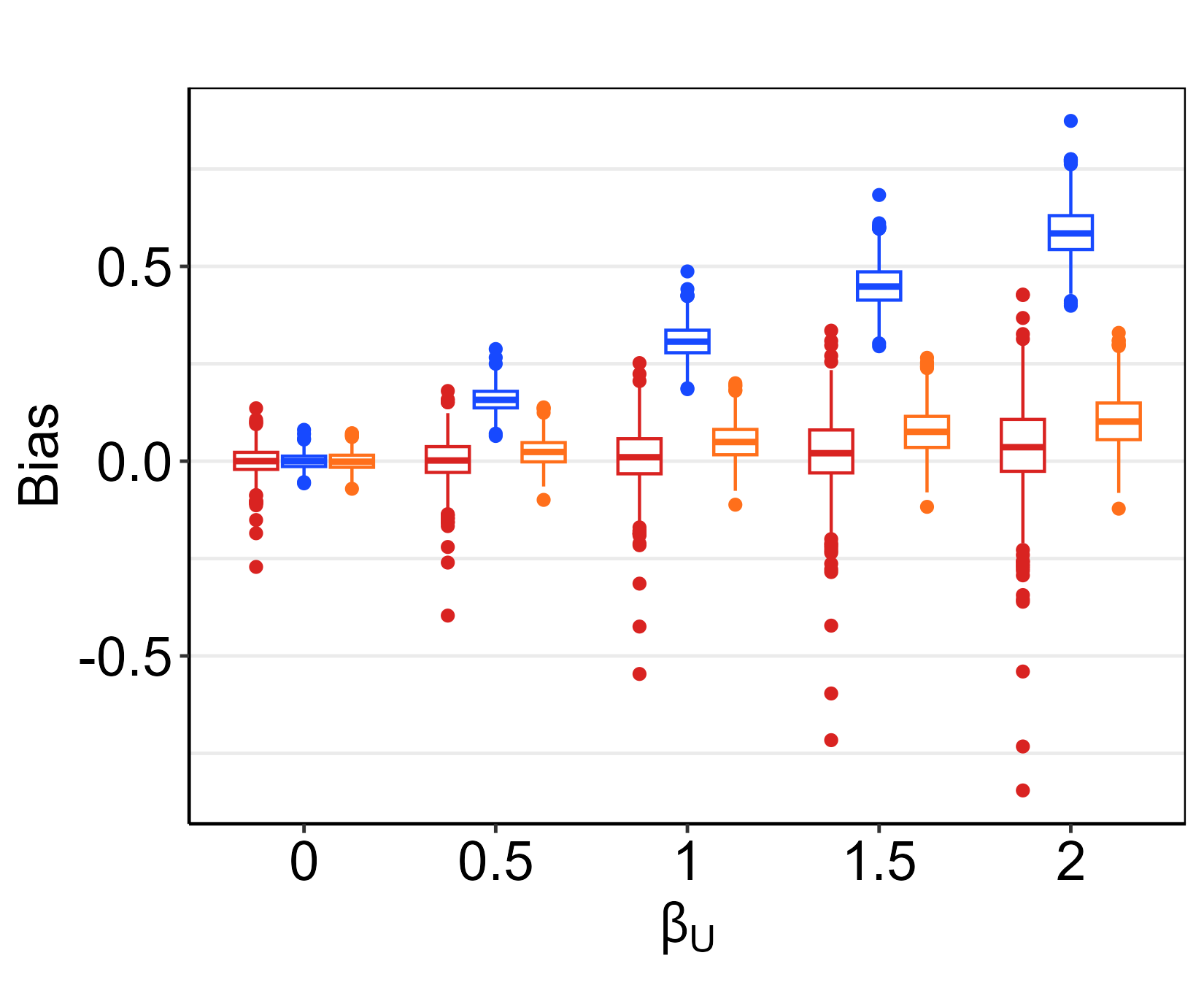}    &  \includegraphics[width = 0.45\textwidth]{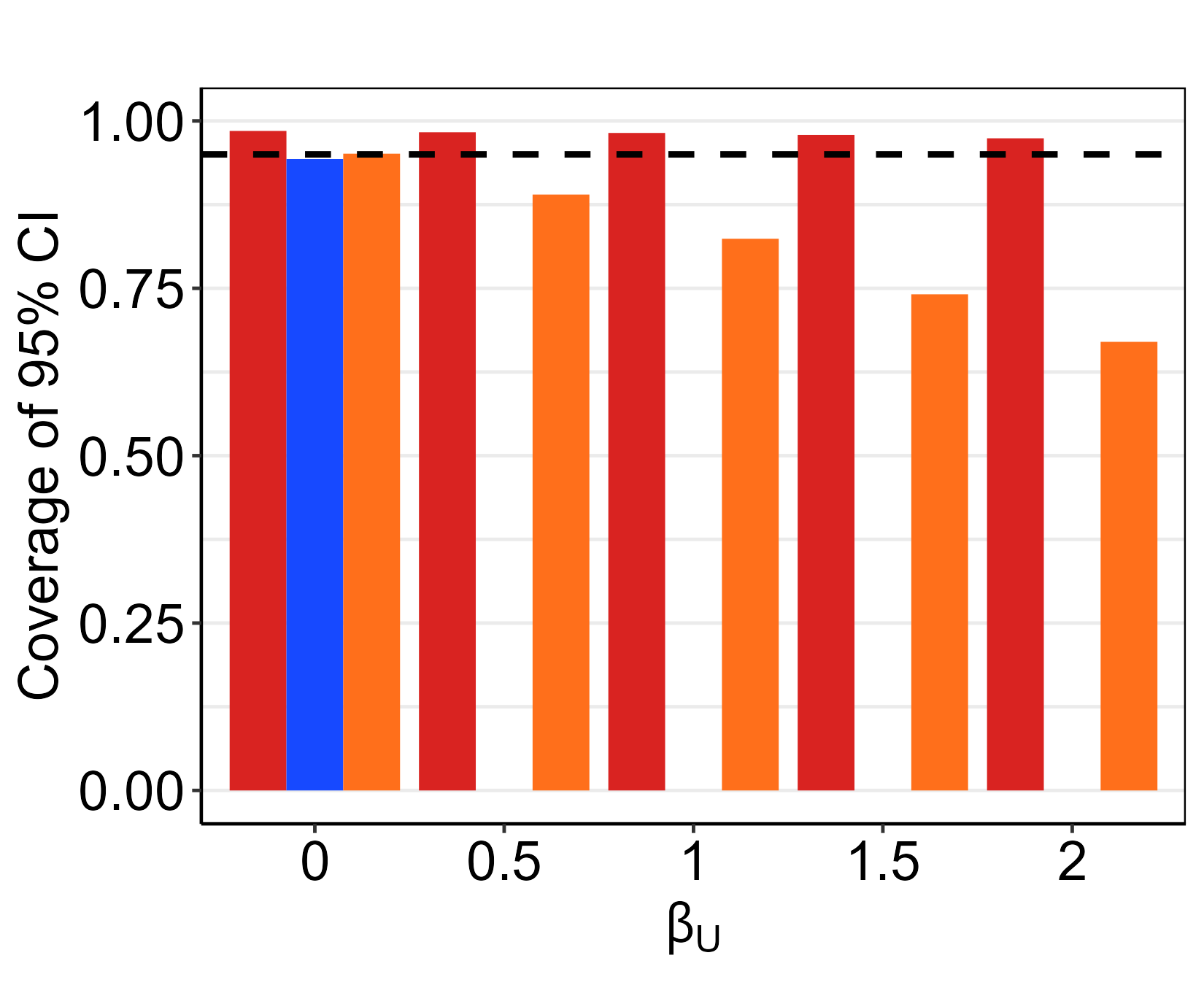} \\
          \includegraphics[width = 0.45\textwidth]{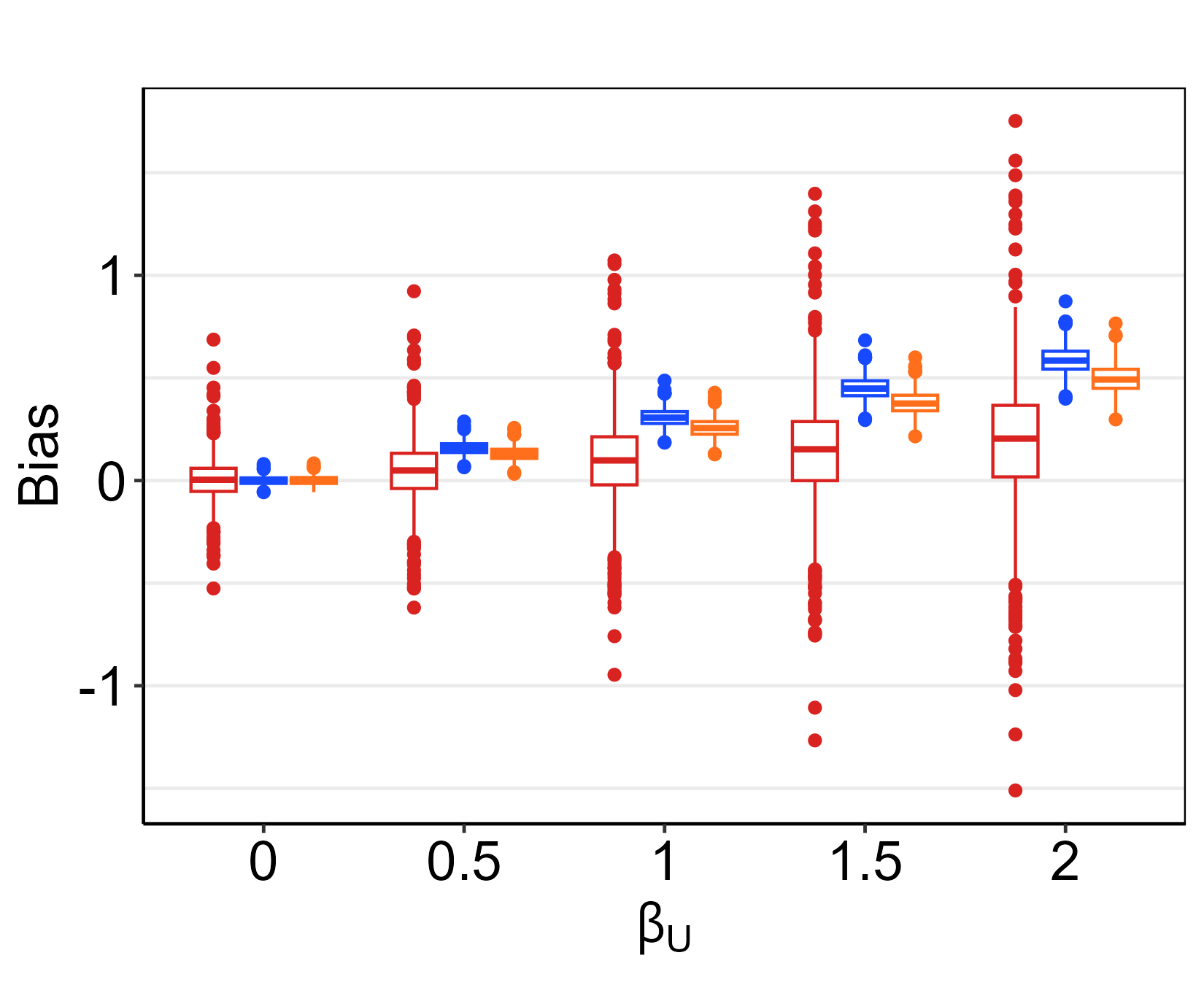}    &  \includegraphics[width = 0.45\textwidth]{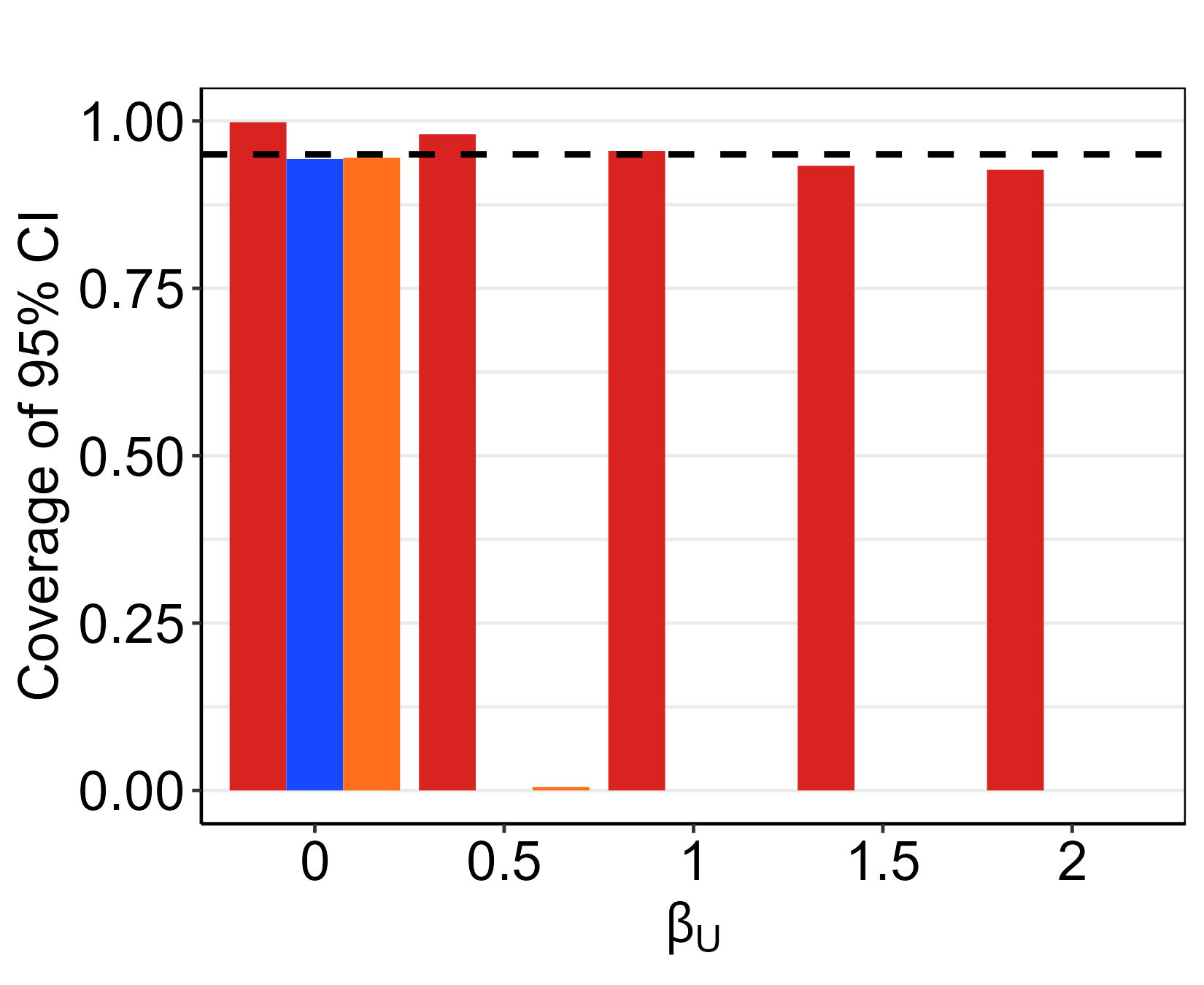} \\
           \includegraphics[width = 0.45\textwidth]{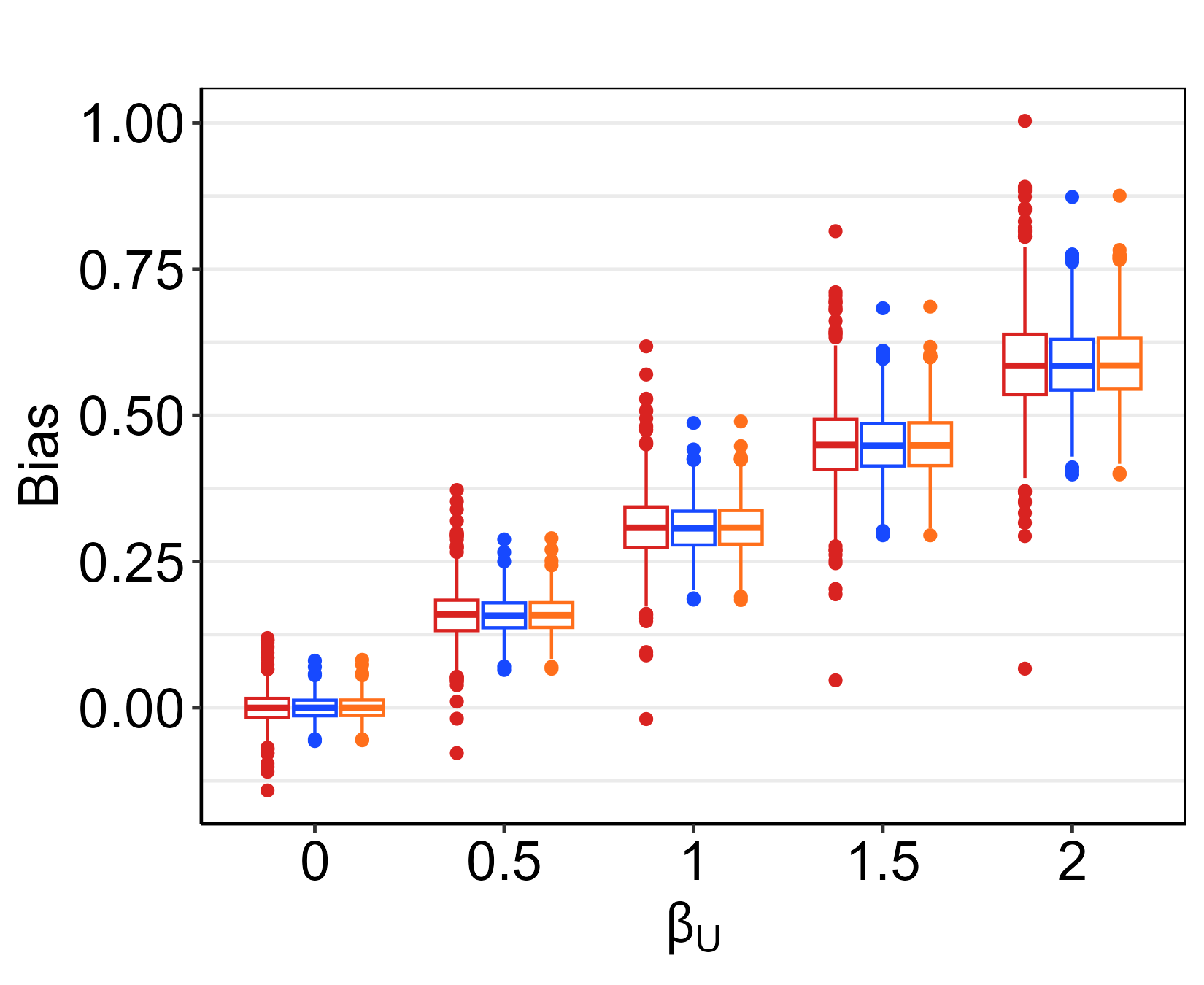}    &  \includegraphics[width = 0.45\textwidth]{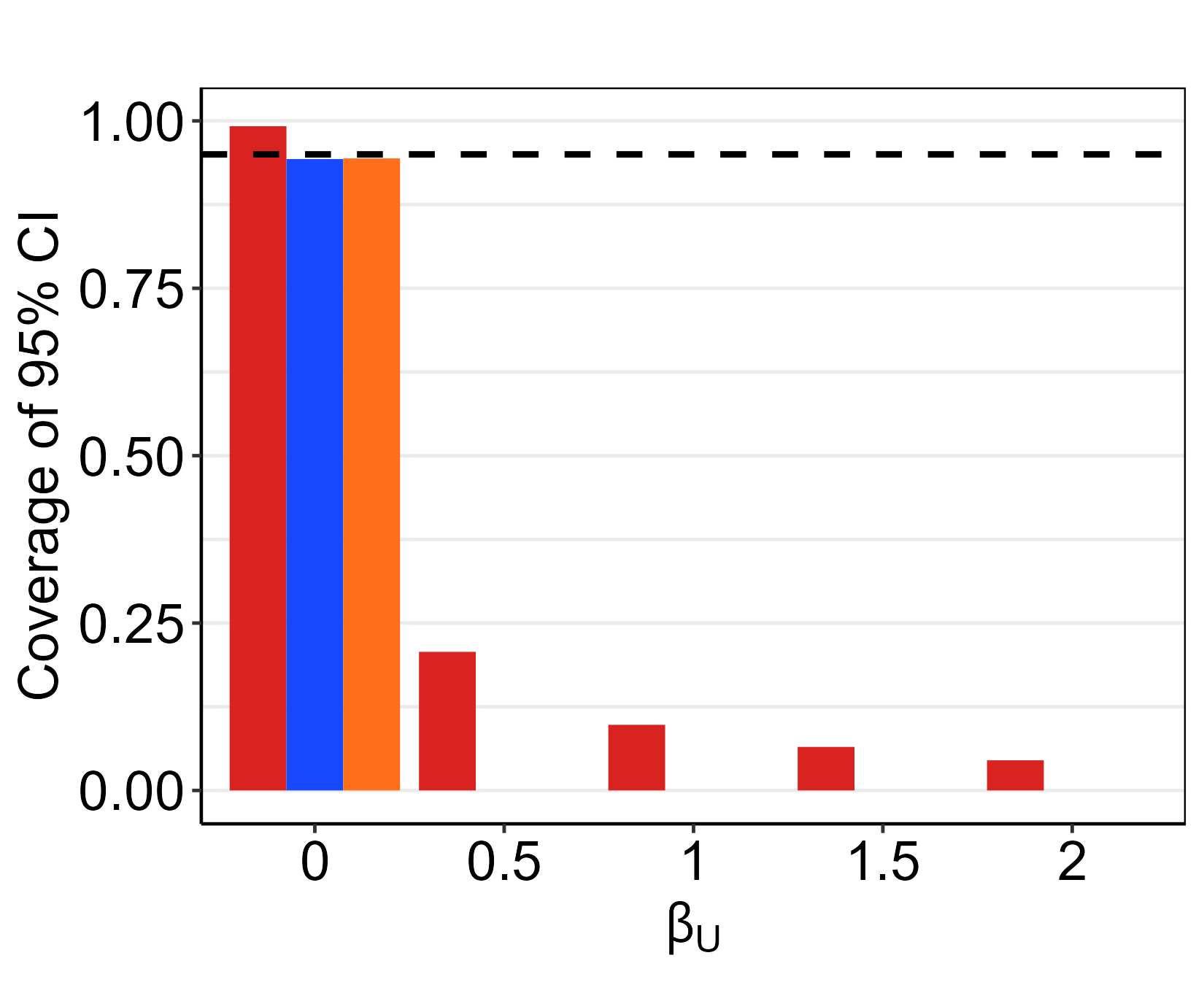} \\
         \multicolumn{2}{c}{\includegraphics[width = 0.7\textwidth]{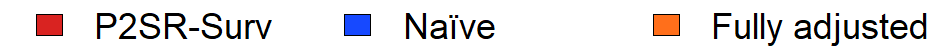}}
        \end{tabular}
        
        \caption{\label{fig:sim}Bias (left) and coverage of 95\% confidence intervals (right) of three methods for $\beta_A$, with $c_U=1$ (top), $0.2$ (middle) or $0$ (bottom).}
    \end{figure}

We report the bias and coverage probabilities of the 95\% confidence intervals for the three methods in Figure~\ref{fig:sim} with different values of $\beta_U$ and $c_U$. When $\beta_U=0$, i.e., there is no unmeasured confounding, all three methods produce unbiased point estimates and their confidence intervals are calibrated. When $c_U=1$, as the unmeasured confounding bias increases, the na\"ive method is severely biased and the coverage probabilities of its 95\% confidence intervals quickly reduce to zero; the fully-adjusted method gives moderately biased estimates and the coverage probabilities of its 95\% confidence intervals reduce to as low as 70\%; P2SR-Surv remains unbiased with calibrated 95\% confidence intervals. We notice that P2SR-Surv produces more variable point estimates than the other two methods, especially with larger unmeasured confounding ($\beta_U$). This is expected -- with large unmeasured confounding, even if the proposed approach is capable of correcting the bias, the loss of efficiency however may be unavoidable.

When $c_U=0.2$, the NC variables have a weaker association with the unmeasured confounders. In this case, the fully-adjusted method produces 95\% confidence intervals with severe undercoverage, whereas the rest of the previous conclusions still hold. The estimators of P2SR-Surv are subject to a notable bias with larger $\beta_U$ and their standard error is even higher, although the 95\% confidence intervals remain calibrated. This phenomenon resembles the issue of ``weak instrument'' in instrumental variable literature -- with proxies less informative about the source of unmeasured confounding, the two-stage regression estimator is biased towards the ``na\"ive'' method estimator, with inflated standard error~\supercite{staiger1994instrumental,andrews2019weak}. Unsurprisingly, when $c_U=0$, all three methods are equally biased in the presence of unmeasured confounding with poor coverage of 95\% confidence intervals -- when the NC variables do not contain any additional information about the unmeasured confounders, it is hopeless to correct the confounding bias. 

\section{Data application: effectiveness of right heart catheterization (RHC) on mortality}\label{sec:rhc}

In this section, we illustrate the proposed approach by comparing mortality among critically ill hospitalized patients who received RHC within first 24 hours of study entry (hospital admission) with those who didn't using data from the Study to Understand Prognoses and Preferences for Outcomes and Risk of Treatments (SUPPORT)~\supercite{connors1996effectiveness,connors1995controlled,connors1996outcomes}. The analysis uses data from 5,735 SUPPORT patients who entered ICU within 24 hours of study entry, with 2,184 receiving RHC in the first 24 hours after study entry as confirmed by chart abstraction and bedside flow sheets.  The dataset includes 71 baseline demographic, clinical and laboratory measurements. The dataset also provides study admission dates, dates of death, and dates of last contact. A total of 3,722 deaths occured during the follow-up, ascertained from clinical records and National Death Index. The median follow-up time is 166 days (Range: 2 days - 5.3 years). For the analysis, we focus on time to date of death since study admission up to 180 days after study entry, during which a total of 2,829 deaths occurred, right-censored at the date of last contact. Following \textcite{tchetgen2020introduction} and \textcite{cui2023semiparametric}, we select  $\text{PaO}_2/\text{FiO}_2$ and $\text{PaCO}_2$ as the NCE $Z$, the blood pH and hematocrit as the NCO $W$, and include the remaining 67 covariates in $X$. For estimation and inference of the adjusted hazards difference of RHC, we implement the proposed two-stage regression method (P2SR-Surv) and three additive hazards regression models: unadjusted, adjusted for $X$ only, and adjusted for $(X, Z, W)$. We summarize the assumptions needed for the proposed two-stage-least-squares method in Appendix S8 and their appropriateness in the RHC example.  From all four methods, RHC is significantly associated with higher mortality, which is consistent with previous findings~\supercite{connors1996effectiveness,cui2023semiparametric}. Compared with the other methods, P2SR-Surv produces a smaller effect size with an adjusted hazard difference of 0.34 per person-year (95\% CI: 0.16, 0.52), consistent with our discussion in Section~\ref{sec:intro} that unmeasured confounding due to disease severity may cause an upward bias of the estimated effect of RHC on mortality in a standard analysis. %ETT. CAN YOU PLEASE ADD NUMBER OF EVENTS OBSERVED EVENTS, FOLLOWUP TIME AND CENSORING RATES OF THE STUDY

\begin{figure}
    \centering
    \includegraphics[width=0.6\linewidth]{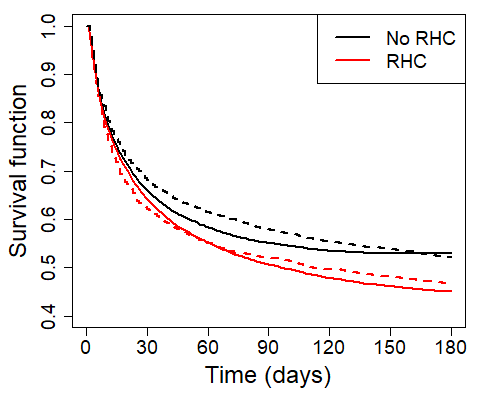}
    \caption{Counterfactual marginal survival curves up to 180 days after hospital admissionusing the proximal inference method in Section ~\ref{sec:survfunc} (solid line)  and inverse treatment probability weighted Kaplan-Meier curves of survival functions (dashed lines) comparing patients with or without RHC.}
    \label{fig:survfunc}
\end{figure}

Figure~\ref{fig:survfunc} reports the counterfactual marginal survival curves based on the proximal inference method in Section~\ref{sec:survfunc} and the standard inverse treatment probability Kaplan-Meier curves, comparing patients with or without receiving RHC.~\supercite{cole2004adjusted} Different from the proposed proximal inference method, the IPW-KM method assume no unmeasured confounding. In this analysis, we estimated the probability of treatment in the IPW-KM method using logistic regression, with all available covariates at baseline $(X,W,Z)$ as indepedent variables. In Figure~\ref{fig:survfunc}, the proximal inference method reports a slightly larger difference in the estimated survival curves during the early follow-up relative to the IPW-KM method, and during the later follow-up the difference appears smaller. The estimated relative risk of death within 30 days using the proximal inference method, defined as the ratio of one minus the survival function at 30 days in the RHC group and that in the no RHC group, is 1.05 (95\% CI: 1.02, 1.08), whereas the relative risk using the IPW-KM method is 1.18 (95\% CI: 1.07, 1.29). The estimated relative risk of death within 180 days using the above two methods are 1.17 (95\% CI: 1.08, 1.26) and 1.12 (95\% CI: 1.04, 1.20), respectively. The 95\% confidence intervals are constructed using nonparametric bootstrap with 2,000 resamples. We note that the counterfactual marginal survival curves in Section~\ref{sec:survfunc} are based on the assumption of constant treatment effect in Equation~\eqref{eq:ah0}. Extension to potential time-varying effects is discussed in Appendix S2.
\begin{table}[!htbp]
    \centering
    \caption{Estimates and 95\% CI of adjusted hazards difference per person-year ($\beta_A$)}
    \label{tab:rhc}
    \begin{tabular}{c|c}
    \hline
           & $\beta_A$\\
           \hline
           P2SR-Surv & $0.34$ ($0.16$, $0.52$)\\
           Unadjusted & $0.40$ ($0.26$, $0.54$)\\
           Adjusted for $X$ & $0.36$ ($0.19$, $0.54$)\\
           Adjusted for $(X, Z, W)$ & $0.37$ ($0.19$, $0.55$)\\
     \hline
    \end{tabular}

    \label{tab:my_label}
\end{table}

%ETT THE PLOT LOOKS GREAT CAN YOU CHANGE A=1 TO A=RHC AND A=0 TO A= NO RHC. ALSO IT DOESNT MAKE SENSE TO PLOT THE UNADJUSTED KAPLAN MEIER CURVES, MAYBE IPTW KM CURVES ADJUSTING FOR MEASURED CONFOUNDERS VIA PROPENSITY SCORE MAKES MORE SENSE GIVEN THAT IS THE STANDARD APPROACH THAN CRUDE ESTIAMTE 

\section{Discussion}\label{sec:discussion}

The proposed two-stage regression approach can be viewed as a specific implementation of the so-called  ``outcome confounding bridge function'' approach in proximal causal inference, which relates the conditional distribution of primary and negative control outcome variables given the other covariates~\supercite{miao2018identifying, tchetgen2020introduction,cui2023semiparametric}. An alternative approach uses the so-called ``treatment confounding bridge function'' that models the inverse propensity score of treatment using the NCE~\supercite{cui2023semiparametric} in case the primary exposure is binary. We leave such developments to future research.% This may be achieved for our approach with a similar development in~\textcite{ying2024proximal} but is likely to come without a straightforward implementation, as indicated in the previous works. Furthermore, our approach accommodates different types of $A$, which may not be easily achieved by a treatment confounding bridge function approach.

Our work has several  limitations. First, as~\textcite{aalen1980model} pointed out, the additive hazards model requires constraints on the parameters such that the resulting hazard function is non-negative. In practice, violation of the constraints may occur with small samples or extreme covariate values, producing a non-monotone survival function. Several post-hoc fixes can be used to resolve this issue. For example, \textcite{aalen1980model} proposed estimating the additive hazards model without restriction and replacing the hazard function during the time when it is negative with zero. Another solution by~\textcite{lin1994semiparametric} was described in Equation \eqref{eq:survcorrect} of Section \ref{sec:survfunc}.  Secondly, the location-shift model in Assumption~\ref{assump:u-model0} requires the error term of the unmeasured confounder to be homoscedastic, that is independent of $(A, Z, X)$, which is technically not empirically testable. In the RHC example, this assumption may fail if recipients of RHC had a wider (or narrower) spread of disease severity.  However, the assumption can be somewhat relaxed, allowing the error distribution to depend on covariates $X$, at the expense of inducing a time-varying effect of $X$ that must be incorporated into the estimated hazard regression models. %THIS DOES NOT SEEM PARTICULARLY PERTINENT.

An R package pci2s for the proposed proximal two-stage regression methods with analytic standard error estimators is available on GitHub at \url{https://github.com/KenLi93/pci2s}. Our method is introduced in the context of right-censored time-to-event data. However, the methods in Sections~\ref{sec:2sls} and \ref{sec:multi-nco} equally apply to left-truncated or interval-censored time-to-event data using additive hazards regression, as proposed by~\textcite{lin1997additive}.  Implementation of these extensions in the R package will be pursued in future work.

\clearpage
\printbibliography

\appendix

\pagebreak

\begin{center}
\textbf{\large Supplementary Material of ``Regression-based proximal causal inference for right-censored time-to-event data''}
\end{center}

\renewcommand\thesection{\Alph{section}}

\setcounter{section}{0}
\renewcommand{\thesection}{S\arabic{section}}

\setcounter{equation}{0}
\renewcommand{\theequation}{S\arabic{equation}}

\setcounter{figure}{0}
\renewcommand{\thefigure}{S\arabic{figure}}

\setcounter{algorithm}{0}
\renewcommand{\thealgorithm}{S\arabic{algorithm}}

\setcounter{assumption}{0}
\renewcommand{\theassumption}{S\arabic{assumption}}

\setcounter{result}{0}
\renewcommand{\theresult}{S\arabic{result}}

\setcounter{table}{0}
\renewcommand{\thetable}{S\arabic{table}}

 \begin{refsection}

\section{Proofs}

Let $A.k$ denote Assumption $k$ for an integer $k$.
\subsection{Proof of Equation (4)}

By Assumption 1, we have
$$\lambda_T(t\mid A, U, X) = \beta_0(t)+\beta_A^TA + \beta_X^TX+\beta_UU.$$

Therefore 
\begin{align}
    P(T>t\mid A, U, X) &= \exp\{-\int_0^t \lambda_T(s\mid A, U, X)ds\}\nonumber\\
    &= \exp\{-\int_0^t\beta_0(s)ds - t(\beta_A^T + \beta_X^TX + \beta_U U)\}\label{eq:ah-surv}
\end{align}
and 
\begin{align*}
    P(T>t\mid A=0, U=0, X=0) &= \exp\{-\int_0^t\beta_0(s)ds \}.
\end{align*}
We thus have
\begin{align*}
    \dfrac{P(T>t\mid A, U, X) }{P(T>t\mid A=0, U=0, X=0)} = \exp\{- t(\beta_A^T + \beta_X^TX + \beta_U U)\}
\end{align*}

\subsection{Proof of Equation (5)}

Under the Consistency condition (1) and Exchangeability condition (2), we have
\begin{align*}
    \lambda_{T^{A=a}}(t\mid U, X)&\stackrel{(2)}{=} \lambda_{T^{A=a}}(t\mid A=a, U, X)\\ &\stackrel{(1)}{=} \lambda_T(t\mid A=a, U, X)\\
    &\stackrel{A.1}{=} \beta_0(t) + \beta_A^Ta + \beta_X^TX+\beta_UU.
\end{align*}

Therefore,
\begin{align*}
    S_{T^{A=a}}(t\mid U, X) &= \exp\left\{-\int_0^t \lambda_{T^{A=a}}(s\mid U, X)\, ds\right\}\\
    &= \exp\left\{-\int_0^t\beta_0(s)ds - t(\beta_A^Ta + \beta_X^TX + \beta_U U)\right\}
\end{align*}
and
\begin{align*}
    S_{T^{A=a}}(t\mid X) &= E\{S_{T^{A=a}}(t\mid U, X)\mid X\}\\
    &= \exp\left\{-\int_0^t\beta_0(s)ds - t(\beta_A^Ta + \beta_X^TX)\right\}E\left\{\exp\left(-t\beta_UU\right)\mid X\right\}.
\end{align*}

The result follows.

\subsection{Proof of Result 1}
By Equation~\ref{eq:ah-surv} and Assumption 2, we therefore have
\begin{align*}
    P(T>t\mid A, U, X, Z) &\stackrel{A.2}{=} P(T>t\mid A, U, X)\\
    &= \exp\{-\int_0^t\beta_0(s)ds - t(\beta_A^TA + \beta_X^TX + \beta_U U)\}\\
    &\stackrel{A.3}{=}\exp[-\int_0^t\beta_0(s)ds - t\{\beta_A^TA + \beta_X^TX + \beta_U E(U\mid A, Z, X)\}-t\beta_U\epsilon]
\end{align*}

On both sides, taking conditional expectation given $(A, X, Z)$, we have
\begin{align*}
    &P(T>t\mid A, X, Z)\\
    &= \exp\left[-\int_0^t\beta_0(s)ds - t\{\beta_A^TA + \beta_X^TX + \beta_U E(U\mid A, Z, X)\}\right]\int \exp\{-t\beta_U\epsilon\}f(\epsilon\mid A, Z, X)d\epsilon\\
    &\stackrel{A.3}{=}\exp\left[-\int_0^t\beta_0(s)ds - t\{\beta_A^TA + \beta_X^TX + \beta_U E(U\mid A, Z, X)\}\right]\int \exp\{-t\beta_U\epsilon\}f(\epsilon)d\epsilon\\
    &= \exp\left[-\int_0^t\beta_0(s)ds+\log \int \exp\{-t\beta_U\epsilon\}f(\epsilon)d\epsilon - t\{\beta_A^TA + \beta_X^TX + \beta_U E(U\mid A, Z, X)\}\right]
\end{align*}

and
\begin{align*}
    \lambda(t\mid A, X, Z) &= -\dfrac{\partial}{\partial t}\log P(T>t\mid A, X, Z)\\
    &= \beta_0(t) - \partial \log\int\exp\{-t\beta_Ue\}dF_\epsilon(e)/\partial t + \beta_A^TA + \beta_X^TX + \beta_UE(U\mid A, Z, X).
\end{align*}

\subsection{Proof of Result 2A}

By Equation~(13) and Assumption 4, we have
\begin{align*}
    E(W \mid A, Z, X) &= c_{01} + c_{U1}(\gamma_0 + \gamma_A^TA + \gamma_Z^TZ + \gamma_X^TX)+c_{X1}^TX\\
    &= (c_{01} + c_{U1}\gamma_0) + c_{U1}\gamma_A^TA + c_{U1}\gamma_Z^TZ + (c_{U1}\gamma_X^T+c_{X1}^T)X\\
    &= c_{01}^* + (c_{A1}^*)^TA + (c_{Z1}^*)^TZ + (c_{X1}^*)^TX.
\end{align*}

Also, by Equation (13), we have
$$E(U\mid A, Z, X)=\{E(W\mid A, Z, X)-c_{01} - c_{X1}^TX\}/c_{U1}.$$

Therefore, by Equation (10) we have
\begin{align*}
    \lambda_T(t\mid A, Z, X) &= \tilde\beta_0(t) + \beta_A^TA + \beta_U\{E(W\mid A, Z, X)-c_{01} - c_{X1}^TX\}/c_{U1} + \beta_X^TX\\
    &= (\tilde\beta_0(t) - \beta_Uc_{01}/c_{U1}) + \beta_A^TA + (\beta_U/c_{U1}) E(W\mid A, Z, X) + (\beta_X - \beta_Uc_{X1}/c_{U1} )^TX\\
    &= \beta_{01}^*(t) + \beta_A^TA + \beta_{U1}^*E(W\mid A, Z, X) + (\beta_{X1}^*)^TX.
\end{align*}

\subsection{Proof of Result 2B}

Given Assumption 6B, taking conditional expectation given $(A, Z, X)$ on both sides of Equation (16), we have
\begin{align*}
    E(W\mid A, Z, X) &= E\{E(W\mid A, U, Z, X)\mid A, Z, X\}\\
    &= E\{\exp(c_{02} + c_{U2}U + c_{X2}^TX)\mid A, Z, X\}.
\end{align*}

By Assumption 3, we have
\begin{align*}
    E(W\mid A, Z, X) &= E\left[\exp\left\{c_{02} + c_{U2}E(U\mid A, Z, X) + c_{X2}^TX + c_{U2}\epsilon\right\}\mid A, Z, X\right]\\
    &= \exp\left\{c_{02} + c_{U2}E(U\mid A, Z, X) + c_{X2}^TX\right\}E\{\exp(c_{U2}\epsilon)\mid A, Z, X\}\\
    &= \exp\left\{c_{02} + c_{U2}E(U\mid A, Z, X) + c_{X2}^TX\right\}E\{\exp(c_{U2}\epsilon)\}\\
    &= \exp\left\{c_{02} + c_{U2}E(U\mid A, Z, X) + c_{X2}^TX\right\}\int \exp(c_{U2}e)dF_\epsilon(e)\\
    &= \exp\left\{c_{02} + \log\int \exp(c_{U2}e)dF_\epsilon(e) + c_{U2}E(U\mid A, Z, X) + c_{X2}^TX\right\}.
\end{align*}

Finally, by Assumption 4, we have
\begin{align*}
    E(W\mid A, Z, X) &= \exp\left\{c_{02} + \log\int \exp(c_{U2}e)dF_\epsilon(e) + c_{U2}(\gamma_0 + \gamma_A^TA + \gamma_Z^TZ + \gamma_X^TX) + c_{X2}^TX\right\}\\
    &= \exp\left\{(c_{02} + \log\int \exp(c_{U2}e)dF_\epsilon(e) + c_{U2}\gamma_0) + c_{U2}\gamma_A^TA +c_{U2}\gamma_Z^TZ + (c_{U2}\gamma_X + c_{X2})^TX\right\}\\
    &= \exp\left\{ c_{02}^{*} + (c_{A2}^*)^T A + (c_{Z2}^*)^T Z + (c_{X2}^*)^TX\right\}.
\end{align*}

On the other hand, let $\mu_2(A, Z, X)$ be the linear predictor of the regression model of $E(W\mid A, Z, X)$, that is,
$$\mu_2(A, Z, X) = c_{02} + \log\int \exp(c_{U2}e)dF_\epsilon(e) + c_{U2}E(U\mid A, Z, X) + c_{X2}^TX.$$
Note that by the above derivation, we also have
$$\mu_2(A, Z, X) =  c_{02}^{*} + (c_{A2}^*)^T A + (c_{Z2}^*)^T Z + (c_{X2}^*)^TX.$$
From Result 1, we have
\begin{align*}
    \lambda_T(t\mid A, Z, X) &= \tilde\beta_0(t) + \beta_A^TA + \beta_U\{\mu_2(A, Z, X) - c_{02} - \log\int \exp(c_{U2}e)dF_\epsilon(e) -  c_{X2}^TX\}/c_{U2} + \beta_X^TX\\
    &= \left[\tilde\beta_0(t)-\beta_U\left\{c_{02} + \log\int \exp(c_{U2}e)dF_\epsilon(e)\right\}/c_{U2}\right] +\\
    &\qquad \beta_A^TA + (\beta_{U2}/c_{U2})\mu_2(A, Z, X) + (\beta_X - \beta_Uc_{X2}/c_{U2})^TX\\
    &= \beta_{02}^{*}(t)+\beta_A^TA + \beta_{U2}^*\mu_2(A, Z, X) + (\beta_{X2}^*)^TX.
\end{align*}

\subsection{Proof of Result 2C}

By Assumption 6C, we have
\begin{align*}
    P(W > t\mid A, U, Z, X) &= \exp\left\{-\int_0^t c_{03}(s)ds - t(c_{U3}U + c_{X3}^TX)\right\}. 
\end{align*}

Taking conditional expectation given $(A,Z,X)$ on both sides, we have
\begin{align*}
    &P(W > t\mid A, Z, X) \\&= E\left\{P(W > t\mid A, U, Z, X)\mid A, Z, X\right\}\\
    &= E\left[\exp\left\{-\int_0^t c_{03}(s)ds - t(c_{U3}U + c_{X3}^TX)\right\}\mid A, Z, X\right]\\
    &= \exp\left[-\int_0^t c_{03}(s)ds - t\left\{c_{U3}E(U\mid A, Z, X) + c_{X3}^TX\right\}\right]E\left\{\exp(-tc_{U3}\epsilon)\mid A, Z, X\right\}\\
    &= \exp\left[-\int_0^t c_{03}(s)ds - t\left\{c_{U3}E(U\mid A, Z, X) + c_{X3}^TX\right\}\right]\int \exp(-tc_{U3}e)dF_\epsilon(e)\\
    &= \exp\left[-\int_0^t c_{03}(s)ds + \log\int \exp(-tc_{U3}e)dF_\epsilon(e)  - t\left\{c_{U3}E(U\mid A, Z, X) + c_{X3}^TX\right\}\right],
\end{align*}

and therefore
\begin{align*}
    &\lambda_W(t\mid A, Z, X)\\ &= -\dfrac{\partial}{\partial t}\log P(W > t\mid A, Z, X) \\
    &= c_{03}(t)-\partial\left\{ \log\int \exp(-tc_{U3}e)dF_\epsilon(e)\right\}/\partial t + c_{U3}E(U\mid A, Z, X) + c_{X3}^TX.
\end{align*}

Under Assumption 4, we further have
\begin{align*}
    &\lambda_W(t\mid A, Z, X) \\&= c_{03}(t)-\partial\left\{ \log\int \exp(-tc_{U3}e)dF_\epsilon(e)\right\}/\partial t + c_{U3}(\gamma_0 + \gamma_A^TA + \gamma_Z^TZ + \gamma_X^TX) + c_{X3}^TX\\
    &= \left[c_{03}(t)-\partial\left\{ \log\int \exp(-tc_{U3}e)dF_\epsilon(e)\right\}/\partial t + c_{U3}\gamma_0\right] + c_{U3}\gamma_A^TA + c_{U3}\gamma_Z^TZ + (c_{U3}\gamma_X + c_{X3})^TX\\
    &= c_{03}^*(t) + (c_{A3}^*)^TA + (c_{Z3}^*)^TZ + (c_X^*)^TX
\end{align*}

On the other hand, let $\mu_3(A, Z, X)$ be the linear predictor in the additive hazard regression model of $\lambda_W(t\mid A, Z, X)$, that is,
\begin{align*}
    \mu_{3}(A, Z, X) 
    &= (c_{A3}^*)^TA + (c_{Z3}^*)^TZ + (c_X^*)^TX\\
    &= c_{U3}E(U\mid A, Z, X) + c_{X3}^TX - c_{U3}\gamma_0.
\end{align*}

By Result 1, we then have 
\begin{align*}
    \lambda_T(t\mid A, Z, X) &= \tilde\beta_0(t) + \beta_A^TA + \beta_U\{\mu_3(A, Z, X)+ c_{U3}\gamma_0 - c_{X3}^TX\}/c_{U3} + \beta_X^TX\\
    &= \{\tilde\beta_0(t) + \beta_U\gamma_0\} + \beta_A^TA + (\beta_{U}/c_{U3})\mu_3(A, Z, X) + (\beta_X - \beta_Uc_{X3}/c_{U3} )^TX\\
    &= \beta_{03}^{*}(t) + \beta_A^TA + (\beta_{U3}^*)^T\mu_3(A, Z, X) + (\beta_{X3}^*)^TX
\end{align*}

\subsection{Proof of Result 3}

By Eq.(7), we have
$$P(T>t\mid A, U, X, Z) = \exp\{-B_0(t)-t\beta_A^TA - t\beta_X^TX - t\beta_U^TU\}.$$

Therefore for each value of $a$, we have
\begin{align*}
    & P(T^{A=a}> t)\\
    =& E\{P(T^{A=a}\mid U, X)\}\\
    =& E\{P(T>t\mid A=a, U, X)\}\qquad\mbox{(by exchangeability and consistency)}\\
    =& E\{P(T>t\mid A=a, U, X, Z)\}\qquad\mbox{(by Assumption 2)}\\
    =& E[\exp\{-B_0(t)-t\beta_A^Ta - t\beta_X^TX - t\beta_U^TU\}]\qquad\mbox{(by Assumption 1)}\\
    =& E[\exp(-t\beta_A^Ta + t\beta_A^T A)\exp\{-B_0(t)-t\beta_A^TA - t\beta_X^TX - t\beta_U^TU\}]\\
    =& E[\exp(-t\beta_A^Ta + t\beta_A^T A)P(T>t\mid A, U, X, Z)]\\
    =& E[\exp(-t\beta_A^Ta + t\beta_A^T A)E\{P(T>t\mid A, U, X, Z)\mid A, X, Z\}]\\
    =& E[\exp(-t\beta_A^Ta + t\beta_A^T A)P(T>t\mid A, X, Z)].
\end{align*}

\newpage
\section{Identification and estimation under general case of Assumption 1*, a nonlinear $E(U\mid A, Z, X)$, and nonlinear models for the NCOs.}

We assume Assumptions 1*, 2, 3 and 5 hold. We remove Assumption 4 and write $E(U\mid A, Z, X)=\gamma(A,Z,X)$ for brevity.

We then have
\begin{align*}
    P(T>t\mid A, U, X, Z) &\stackrel{A.2}{=}P(T>t\mid A, U,X)\\
    &= \exp\left\{-\int_0^t\lambda_T(u\mid A, U, X)du\right\}\\
    &= \exp\left\{-B_0(t) - B_{A,X}(t,A,X)- B_U(t)^TU\right\}
\end{align*}
where $B_0(t, X)=\int_0^t\beta_0(r, X)dr$, $B_{A,X}(t,A,X)=\int_0^t\beta_{A,X}(r,A,X)dr$, and $B_U(t)=\int_0^t\beta_U(r)dr$.

Under Assumption 3, we have:
\begin{align*}
    P(T>t\mid A, U, X, Z) &= \exp\{-B_0(t,X)-B_{A,X}(t,A, X) - B_U(t)\gamma(A, Z,X) - B_U(t)\epsilon\}.
\end{align*}

Integrating over the conditional distribution of $U$ given $(A,Z,X)$ on both sides, we have
\begin{align*}
    P(T>t\mid A, X, Z) &= \exp\{-B_0(t,X)-B_{A,X}(t,A,X) -B_U(t)\gamma(A,Z,X)\}\int \exp\{-B_U(t)e\}f_\epsilon(e)de
\end{align*}
which implies an additive hazards model
\begin{equation}\label{eq:nco-lin-2s}
    \lambda_T(t\mid A, Z,X)=\exp\{\widetilde\beta_0(t, X) + \beta_{A,X}(t,A,X)+\beta_U(t)\gamma(A,Z,X)\}
\end{equation}
where $\widetilde\beta_0(t, X)=\beta_0(t, X) - \partial[\log\int\exp\{-B_U(t)e\}f_\epsilon(e)de]/\partial t$.

We further extend Assumption 6A to Assumption~ below. Extension of Assumptions 6B and 6C are omitted but can be easily derived from the exposition.

\begin{assumption}[NCO follows a generalized linear additive model]\label{assump:w-linear-2}
    \begin{equation}\label{eq:w-linear-2}E(W\mid A, U, Z, X)=c_{X}(X) + c_{U}U,\end{equation}
    where $c_{X}$ is an unknown function.
\end{assumption}

Taking expectation with respect to the conditional distribution of $U$ given $(A,Z,X)$ on both sides of Equation~\eqref{eq:w-linear-2}, under Assumption 3, we have
\begin{align}
    E(W\mid A, Z, X)=c_{X}(X)+c_U^T\gamma(A,Z,X) \label{eq:nco-lin-1s}
\end{align}
where $\gamma(0, 0, X)=0$.

Suppose the functions in Equations~\eqref{eq:nco-lin-2s} and \eqref{eq:nco-lin-1s}, including $\beta_{A,X}(t,A,X)$, $\beta_U(t)$, $\gamma(A,Z, X)$, and $c_X(X)$, follow parametric models with unknown finite-dimensional parameters, i.e. $\beta_{A,X}(t,A,X)=\beta_{A,X}(t,A,X;\xi_1)$, $\beta_U(t)=\beta_U(t;
\xi_2)$, $\gamma(A, Z, X)=\gamma(A, Z, X;\xi_3)$, and $c_X(X)=c_X(X;\xi_4)$. Then the parameters $\xi_1$, \dots, $\xi_4$ may be estimated based on Equations~\eqref{eq:nco-lin-2s} and \eqref{eq:nco-lin-1s} using a two-stage generalized method-of-moments approach similar to Algorithm 1. To ground ideas, in Stage 1, we estimate the parameters $c_U$, $\xi_3$ and $\xi_4$ by solving the estimating equation
$$\sum_{i=1}^n g_1(A_i,Z_i,X_i)[W_i - c_X(X_i;\xi_4)-c_U\gamma(A_i,Z_i,X_i;\xi_3)]=0$$
where $g_1(A_i,Z_i,X_i)$ is a vector-valued user-specified function with dimension larger than the sum of dimensions of $\xi_3$, $\xi_4$ and $c_U$. Denoting the estimators as $\widehat\xi_3$, $\widehat\xi_4$ and $\widehat c_U$, in Stage 2 we may estimate $\xi_1$ and $\xi_2$ by solving the estimating equation
\begin{align*}&\sum_{i=1}^n \int g_2(t, A_i,Z_i,X_i)\big[dN_i(t)- R_i(t)d\Lambda_0(t;\xi_1,\xi_2) -\\&\qquad \{\beta_{A,X}(A_i,X_i;\xi_1)+\beta_U(t;\xi_2)\gamma(A_i,Z_i,X_i;\widehat\xi_3)\}R_i(t)dt\big]= 0\end{align*} 
where $g_2(t,A_i,Z_i,X_i)$ is a vector-valued user-specified function with dimension larger than the sum of dimensions of $\xi_1$ and $\xi_2$ and 
$$\Lambda_0(t; \xi_1, \xi_2) = \int_0^t \dfrac{\sum_{j=1}^n\left[dN_j(u) - R_j(u)\{\beta_{A,X}(u, A_j, X_j;\xi_1) + \beta_U(t;\xi_2)\gamma(A_j, Z_j,X_j;\widehat\xi_3)\}du\right]}{\sum_{j=1}^n R_j(u)}.$$

This encompasses the scenario discussed in Section 2.1.

Alternatively, the nonlinear functions can also be modeled using nonparametric methods, such as regression splines~\supercite{wood2003thin} or functions in a Reproducible Kernel Hilbert Space~\supercite{rosipal2001kernel}. We may also consider the model that $\beta_{A,X}(t, A, X)=\beta_1(t)A + \beta_2(t)AX$, so that time-varying effects can be estimated nonparametrically.

Finally, similar to Equation (4), under Assumption 1* and the additional Consistency and Exchangeability assumptions, we have that
$$ \dfrac{S_{T^{A=1}}(t\mid X)}{S_{T^{A=0}}(t\mid X)} = \exp\left(-\{B_{A,X}(t,1,X)\}\right).$$

\newpage
\section{Estimating the standard error of  $\widehat\beta_A$}

In this section, we derived the asymptotic variance of $\widehat\beta_A$, the estimator of $\beta_A$ by Algorithm 1. Let $0\equiv T_{(0)} < T_{(1)}<\dots<T_{(d)}$ be the ordered event times and $d_i$ be the order of $T_i$. That is, we have $T_i = T_{(d_i)}$. Write $N_i(t)=\mathbbm 1(T_i\leq t, \Delta_i=1)$ as the counting for the $i$th subject, and $R_i(t)=\mathbbm 1(T_i\geq t, C_i\geq t)$ as the at-risk process for the $i$th subject.

Suppose in Step 1, we fit an ordinary least square regression model, i.e. we estimate the regression coefficients $c_1^* = (c_{01}^*, (c_{A1}^*)^T, (c_{Z1}^*)^T, (c_{X1}^*)^T)^T$ by solving the estimating equation
$$\sum_{i=1}^n G_{1i}(c_1^*)\equiv\sum_{i=1}^n \begin{pmatrix}
    1 \\ A_i \\ Z_i \\ X_i
\end{pmatrix}\{W_i -\mu_1(A_i, Z_i, X_i;c_1^*)\}=0.$$
where
$$\mu_1(A, Z, X;c_1^*)= c_{01}^* + (c_{A1}^*)^TA + ( c_{Z1}^*)^TZ + (c_{X1}^*)^TX.$$

Further suppose in Step 3, we fit the semiparametric additive model proposed by~\textcite{lin1994semiparametric} with predictors $A$, $\mu_1(A, Z, X;\widehat c_1^*)$ and $X$, which solves the estimating equation
$$\sum_{i=1}^nG_{2i}(c_1^*,\beta_1^*) \equiv \sum_{i=1}^n \int \{S_{2i}(c_1^*)-\overline S_{2}(t,c_1^*)\}\{dN_i(t)- R_i(t)d\hat \Lambda_0(t, c_1^*, \beta_1^*) -(\beta_1^*)^TS_{2i}(c_1^*)R_i(t)dt\} = 0$$where 
\begin{align*}
    \beta_1^* &= (\beta_A^T, (\beta_{U1}^*)^T, (\beta_{X1}^*)^T)^T,\\
    S_{2i}(c_1^*) &= (A_i^T, \mu_1(A_i, Z_i, X_i;  c_1^*)^T, X_i),\\
    \overline S_{2}(t, c_1^*) &= \sum_{j=1}^n S_{2j}(c_1^*)R_j(t) /\sum_{j=1}^n R_j(t),\\
\end{align*}
and

$$\hat \Lambda_0(t, c_1^*,\beta_1^*) = \int_0^t \dfrac{\sum_{j=1}^n\{dN_j(u) - R_j(u)(\beta_{1}^*)^T S_{2i}(c_1^*)du\}}{\sum_{j=1}^n R_j(u)}.$$

Write the vector of unknown parameters as $\theta_1^* = ((c_1^*)^T, (\beta_1^*)^T)^T$, then the above two above two-step algorithm is equivalent to solving the estimating equation

\begin{equation}\label{eq:gmm-linear}\sum_{i=1}^n G_i(\theta_1^*) = \sum_{i=1}^n \begin{pmatrix}
    G_{1i}(c_1^*)\\
    G_{2i}(c_1^*,\beta_1^*)
\end{pmatrix}=0.\end{equation}

By Theorem 3.1 of \textcite{hansen1982large}, under mild regularity conditions, the solution to Equation~\ref{eq:gmm-linear} $\widehat \theta_1^* = ((\widehat c_1^*)^T, (\widehat \beta_1^*)^T)^T$ is asymptotically normal and its variance-covariance matrix can be approximated by the ``sandwich'' estimator
$$\dfrac{1}{n}\{D(\widehat \theta_1^*)\}^{-1}V(\widehat \theta_1^*)\{D(\widehat \theta_1^*)\}^{-T},$$
where
\begin{align*}
    D(\theta_1^*) = \dfrac{1}{n}\sum_{i=1}^n\begin{pmatrix}
        \dfrac{\partial}{\partial (c_1^*)^T} G_{1i}(c_1^*) & 0 \\
        \dfrac{\partial}{\partial (c_1^*)^T} G_{2i}(c_1^*,\beta_1^*) & \dfrac{\partial}{\partial (\beta_1^*)^T} G_{2i}(c_1^*,\beta_1^*)
    \end{pmatrix}
\end{align*}
and
$$V(\theta_1^*) = \dfrac{1}{n}\sum_{i=1}^n\begin{pmatrix}
         G_{1i}(c_1^*)G_{1i}(c_1^*)^T &  G_{1i}(c_1^*)G_{2i}(c_1^*,\beta_1^*)^T \\
         G_{2i}(c_1^*,\beta_1^*)G_{1i}(c_1^*)^T &  G_{2i}(c_1^*,\beta_1^*)G_{2i}(c_1^*,\beta_1^*)^T\end{pmatrix}.$$

Derivation for the approximate variances of estimators by Algorithms 2 and 3 are similar, except that the estimating function $G_{1i}$ is replaced with the corresponding estimating function to the regression model used. For Algorithm 4, the estimating function $G_{1i}$ is replaced by stacking together all estimating functions for parameters indexing the regression models for each of the NCOs.

\newpage
\section{Competing risks as negative controls}\label{sec:cmprsk}

In this appendix, we will show that in the presence of with competing risks, if the following conditions satisfy that (1) the cause-specific hazards functions of the competing risks do not depend on the primary exposure, and (2) the cause-specific hazards function of the primary event does not depend on the negative control exposure, time to the competing risks may serve as a negative control outcome. Therefore, a slightly modified Algorithm 3 may be used for unmeasured confounding bias adjustment. Mathematical proofs of results in this section are deferred to Section \ref{sec:proof-cmprsk}.

For demonstration purposes, we only consider one competing risk event, but extension to multiple competing risks is straightforward. We let $A$ be the primary exposure of interest, $T_0$ be the (uncensored) time to a primary event of interest, $X$ be the measured baseline covariates, $U$ be the unmeasured baseline confounders, $Z$ be the negative control exposure (NCE), $T_1$ be the (uncensored) time to a competing risk, and $T=\min(T_0, T_1)$ be the (uncensored) time to either event. Let $J$ be the cause of event such that $J=j$ if $T=T_j$, $j=0,1$. Let $C$ be the right censoring time due to other reasons. 

Similar to Section 2.3, we make the following assumptions:

\begin{assumption}[Cause-specific hazards function of the primary event]\label{assump:cprisk-ah}
    The conditional cause-specific hazard function for $T_0$ given $(A, U, X, Z)$ is
     \begin{equation}\label{eq:cshaz-primary}\lambda_0(t\mid A, U, X, Z) \equiv \lim_{d t\rightarrow 0}\dfrac{P(t\leq T < t + dt, J = 0\mid A, U, X, Z)}{dt}= \beta_{00}(t) + \beta_A^T A + \beta_{0X}^TX + \beta_{0U} U.\end{equation}
\end{assumption}
Above, the parameter $\beta_A$ encodes the effect of the exposure $A$ on the primary event of interest. The right-hand side of Eq.~\eqref{eq:cshaz-primary} does not depend on $Z$. A sufficient condition for this is $(T_0, T_1)\indep Z$. 

\begin{assumption}[Cause-specific hazards function of the competing risk]\label{assump:nco-ah}
\begin{equation}\label{eq:cshaz-cmprsk}\lambda_1(t\mid A, U, X, Z) \equiv \lim_{d t\rightarrow 0}\dfrac{P(t\leq T < t + dt, J = 1\mid A, U, X, Z)}{dt}= \beta_{10}(t) + \beta_{1X}^TX + \beta_{1U}^T U .\end{equation}
\end{assumption}
The right-hand side of Eq.~\eqref{eq:cshaz-cmprsk} does not depend on $(A, Z)$, analogous to the requirement of conditional independence for an NCO.

Similar to Assumption 3, we require the following assumption of conditional independence censoring so that the cause-specific hazard functions can be identified in the presence of right censoring.
\begin{assumption}[Conditional independence censoring]\label{assum:nc}
    $C\indep(T_0,T_1)\mid A, X, Z$.
\end{assumption}

Similar to Section 3, additional assumptions on the unmeasured confounding $U$ are needed for bias correction
\begin{assumption}(Location-shift model for $U$)\label{assump:u-model}
    \begin{equation}\label{eq:u-model}U=\gamma(A, Z, X)+\epsilon\end{equation} where $\gamma(A, Z, X) = E(U\mid A, Z, X)$, $E(\epsilon)=0$ and $\epsilon\indep (A, Z, X)$.
\end{assumption}

\begin{assumption}[Linear mean model for $U$]\label{assump:u-lm}
\begin{equation}\label{eq:u-lm}
 E(U\mid A, Z, X):=\gamma_{0}+\gamma_{A}^TA + \gamma_{Z}^TZ + \gamma_{X}^TX.
\end{equation}
\end{assumption}

 With derivation similar to that in Section 2, we can show that the cause-specific hazard functions conditioning on the observed $(A, Z, X)$ are

\begin{result}\label{rslt:cprisk-2sls}
Under Assumptions~\ref{assump:cprisk-ah}-\ref{assump:u-lm}, we have
\begin{align}
    \lambda_1(t\mid A, Z, X)&= \beta_{10}^*(t) + (\beta_{1A}^*)^TA + (\beta_{1Z}^*)^T Z +(\beta_{1X}^*)^TX\label{eq:id-cprisk-t1} \\
    \lambda_0(t\mid A, Z, X)&= \beta_{00}^*(t) + \beta_A^TA +(\beta_{0X}^*)^TX + \beta_{0U}^* \mu_1(A, Z, X)\label{eq:id-cprisk-t0} 
\end{align}

where $\beta_{10}^*(t)= \beta_{10}(t) + E[\beta_{1U}\epsilon\exp\{-(\beta_{1U}+\beta_{0U})\epsilon t\}]/E[\exp\{-(\beta_{1U}+\beta_{0U})\epsilon t\}] + \beta_{1U}\gamma_0$, $\beta_{1A}^*=\gamma_{A}^T\beta_{1U}$, $\beta_{1Z}^*=\gamma_Z^T\beta_{1U}$, $\beta_{1X}^*=\gamma_U^T\beta_{1U} + \beta_{1X}$, $\beta_{00}^*(t) =\beta_{00}(t) + E[\beta_{0U}\epsilon\exp\{-(\beta_{1U}+\beta_{0U})\epsilon t\}]/E[\exp\{-(\beta_{1U}+\beta_{0U})\epsilon t\}] - \beta_{0U}\gamma_0$, $\beta_{0U}^* = \beta_{0U}/\beta_{1U}$, $\beta_{1X}^*=\beta_{0X} - \beta_{1X}\beta_{0U}/\beta_{1U}$, and 
\begin{align*}
\mu_1(A, Z, X) &= (\beta_{1A}^*)^T A + (\beta_{1Z}^*)^T Z + (\beta_{1X}^*)^TX.
\end{align*}
\end{result}

The regression coefficients in Equations~\eqref{eq:id-cprisk-t1} and \eqref{eq:id-cprisk-t0} can be identified using cause-specific hazard regression with additive hazards models~\supercite{austin2016introduction}. We summarize the estimation in Algorithm~\ref{alg:p2sls-surv-cprisk}.

\begin{algorithm}[!h]
    \caption{P2SR-Surv with a competing event as an NCO}
    \label{alg:p2sls-surv-cprisk}
    \begin{algorithmic}[1]
        \State Fit a cause-specific additive hazards regression model according to Equation~\eqref{eq:id-cprisk-t1}  and obtain the estimators for the regression coefficients $\widehat\beta_{1A}^*$, $\widehat\beta_{1Z}^*$ and $\widehat\beta_{1X}^*$;
        \State  Obtain the linear predictors $$\widehat\mu(A, Z, X)=(\widehat \beta_{1A}^*)^TA + (\widehat \beta_{1Z}^*)^TZ + (\widehat \beta_{1X}^*)^TX.$$
        \State Fit the cause-specific additive hazards regression model according to Equation~\eqref{eq:id-cprisk-t0} with $\mu(A, Z, X)$ replaced by $\widehat\mu(A, Z, X)$. The regression coefficient for $A$ is an estimator of $\beta_A$.
    \end{algorithmic}
\end{algorithm}

Similar to Result 3, marginal counterfactual cumulative incidence function can be identified and therefore estimated. We present the result below:

\begin{result}\label{rslt:cif}
    Under Assumptions~\ref{assump:cprisk-ah}, \ref{assump:nco-ah}, \ref{assump:u-model} and the additional assumptions of consistency, exchangeability and positivity, the counterfactual marginal survival function is
    \begin{align*}
S_{T^{(a)}}(t) &:= P(T^{(a)}\geq t) \\
&= E[P(T\geq t\mid A=a, Z, X)],
    \end{align*}
    and, for $j=0,1$, the counterfactual marginal cause-specific hazard functions are
    \begin{align*}
        \lambda_{T_j^{(a)}}(t) &=\dfrac{E[\lambda_j(t\mid A=a, Z, X)P(T\geq t\mid A=a, Z, X)}{E[P(T\geq t\mid A=a, Z, X)]}.
    \end{align*}
\end{result}

The counterfactual marginal cumulative incidence functions can further be estimated as
$$F_j^{(a)}(t) = \int_0^t \lambda_{T_j^{(a)}}(r)S_{T^{(a)}}(r)dr.$$

Finally, the above results can further be extended to the case where the completing risk is a valid NCO only after an initial period. That is, the cause-specific hazard function of the competing risk may depend on $A$ during an initial period $(0,D]$ but not after time $D$. We formalize this as Assumption~\ref{assump:nc-period-ah} below: 
\begin{assumption}[Cause-specific hazards function of the competing risk]\label{assump:nc-period-ah}
\begin{equation}\label{eq:cshaz-cmprsk}\lambda_1(t\mid A, U, X, Z) \equiv \lim_{d t\rightarrow 0}\dfrac{P(t\leq T < t + dt, J = 1\mid A, U, X, Z)}{dt}= \beta_{10}(t) + \beta_{1A}(t)^TA + \beta_{1X}(t)^TX + \beta_{1U}(t) U .\end{equation}
where \begin{align*}\beta_{1A}(t) = \begin{cases}b_{A} & \text{if $t \leq D$}\\ 0 &\text{if $t > D$} \end{cases} \quad \beta_{1X}(t) = \begin{cases}b_{X1} & \text{if $t \leq D$}\\ b_{X2} &\text{if $t > D$} \end{cases} \quad  \beta_{1U}(t) = \begin{cases}b_{U1} & \text{if $t \leq D$}\\ b_{U2} &\text{if $t > D$} \end{cases}\end{align*}
\end{assumption}

As before, we could identify $\beta_A$ based on the following result:
\begin{result}\label{rslt:cmprisk-nc-period}
    Under Assumptions~\ref{assump:cprisk-ah}, \ref{assump:u-model}, \ref{assump:u-lm}, and \ref{assump:nc-period-ah}, we have
\begin{align}
        \lambda_1(t\mid A, X, Z) &= b_{12}^*(t) + (b_{A2}^*)^T A + (b_{X2}^*)^T X + (b_{Z2}^*)^T Z, \quad t>D\label{eq:first-stage}\\
        \lambda_1(t\mid A, X, Z) &= b_{11}^*(t) + (b_{A1}^*)^T A + (b_{X1}^*)^T X + b_{U1}^*\mu(A, X, Z), \quad t\leq D\label{eq:anx-model}\\
        \lambda_0(t\mid A, X, Z) &= \beta_{00}^*(t) + \beta_{A}^TA + (\beta_{0X}^*)^TX + \beta_{0U}^* \mu(A, X, Z) \label{eq:second-stage}
\end{align}

where  $b_{12}^*(t) = \widetilde\beta_{10}(t) + b_{U2}\gamma_0$, $b_{A2}^* = b_{U2}\gamma_A$, $b_{X2}^*=b_{X2}+b_{U2}\gamma_X$, $b_{Z2}^* = b_{U2}\gamma_Z$,  $b_{11}^*(t) = \widetilde\beta_{11}(t)$, $b_{A1}^*=b_A$, $b_{X1}^*=b_{X1}-b_{U1}b_{X2}/b_{U2}$, $b_{U1}^*=b_{U1}/b_{U2}$, $\beta_{00}^*(t) = \widetilde \beta_{00}(t)$, $\beta_{0X}^* = \beta_{0X}-\beta_{0U}b_{X2}/b_{U2}$,  $\beta_{0U}^*=\beta_{0U}/b_{U2}$, and
 $$\mu(A, X, Z) = (b_{A2}^*)^T A + (b_{X2}^*)^T X + (b_{Z2}^*)^T Z.$$

Under  the additional assumptions of consistency, exchangeability and positivity, the counterfactual marginal survival function is
    \begin{align*}
S_{T^{(a)}}(t) &:=E\bigg[\exp\{-B_A(t)^Ta + B_A(t)^TA \}P(T> t\mid A, Z, X)\bigg],
    \end{align*}
    and, the counterfactual marginal cause-specific hazard function for the primary event of interest is
    \begin{align*}\lambda^a_0(t)& = \widetilde \beta_{00}(t) + \beta_A^Ta + \dfrac{E[\{\beta_{0X}^TX + \beta_{0U}\gamma(A, Z, X)\}\exp\{-B_A(t)^Ta + B_A(t)^TA \}P(T\geq t\mid A, Z, X)]}{E[\exp\{-B_A(t)^Ta + B_A(t)^TA \}P(T\geq t\mid A, Z, X)]}\\
&=\dfrac{E[\{\lambda_0(t\mid A, Z, X)+\beta_A^Ta -\beta_A^TA \}\exp\{-B_A(t)^Ta + B_A(t)^TA \}P(T\geq t\mid A, Z, X)]}{E[\exp\{-B_A(t)^Ta + B_A(t)^TA \}P(T\geq t\mid A, Z, X)]}
\end{align*}

\end{result}

\newpage
\section{Proof of results in Section~\ref{sec:cmprsk}}\label{sec:proof-cmprsk}
\subsection*{Proof of Result~\ref{rslt:cprisk-2sls}}

 Note that Eq.~\eqref{eq:cshaz-primary} is equivalent to
    \begin{align*}
        E\{dN_0(t)\mid T\geq t, A, U, X, Z\} = \{\beta_{00}(t) + \beta_A^TA + \beta_{0X}^TX + \beta_{0U}\gamma(A, X, Z)+\beta_{0U}\epsilon\},
    \end{align*}
    and therefore
    \begin{align*}
    &E\{dN_0(t)\mid T\geq t, A,  X, Z\}\\
    = & E\left[E\{dN_0(t)\mid T\geq t, A, U, X, Z\}\mid T\geq t,A, X, Z\right] \\ 
    = & E[\{\beta_{00}(t) + \beta_A^TA + \beta_{0X}^TX + \beta_{0U}U\}\mid T\geq t, A, X, Z]\\
    =& \{\beta_{00}(t) + \beta_A^TA + \beta_{0X}^TX\} + E\{\beta_{0U}U \mid T\geq t,A, X, Z\}\\
    =& \{\beta_{00}(t) + \beta_A^TA + \beta_{0X}^TX \} + \int \beta_{0U}u f(u\mid T\geq t, A, X, Z)du\\
    =& \{\beta_{00}(t) + \beta_A^TA + \beta_{0X}^TX \} + \int \beta_{0U}u S(t\mid A, U=u, X, Z)f(u\mid A, X, Z)du / S(t\mid A, X, Z)\\
    =&\qquad \{\beta_{00}(t) + \beta_A^TA + \beta_{0X}^TX + \beta_{0U}\gamma(A, X, Z)\}dt + \dfrac{\int \beta_{0U}\epsilon \exp\{-(\beta_{0U}+\beta_{1U})^T\epsilon t\}f(\epsilon) d\epsilon }{\int \exp\{-(\beta_{0U}+\beta_{1U})^T\epsilon t\}f(\epsilon) d\epsilon}dt
    \end{align*}

    Therefore \begin{align*}
        \lambda_0(t\mid A, X, Z) &= \{\beta_{00}(t) + \beta_A^TA + \beta_{0X}^TX + \beta_{0U}\gamma(A, X, Z)\} + \dfrac{\int \beta_{0U}\epsilon \exp\{-(\beta_{0U}+\beta_{1U})^T\epsilon t\} f(\epsilon)d\epsilon }{\int \exp\{-(\beta_{0U}+\beta_{1U})^T\epsilon t\}f(\epsilon) d\epsilon}\\
        &= \overline\beta_{00}(t) + \beta_A^TA + \beta_{0X}^TX + \beta_{0U}\gamma(A, X, Z)
    \end{align*}
    where $$\overline\beta_{00}(t) = \beta_{00}(t) + \dfrac{\int \beta_{0U}\epsilon \exp\{-(\beta_{0U}+\beta_{1U})^T\epsilon t\}f(\epsilon) d\epsilon }{\int \exp\{-(\beta_{0U}+\beta_{1U})^T\epsilon t\}f(\epsilon) d\epsilon}.$$
    Similarly
    \begin{align*}
        \lambda_1(t\mid A, X, Z) = \overline\beta_{10}(t) + \beta_{1X}^TX + \beta_{1U}\gamma(A, X, Z)
    \end{align*}
    where $$\overline\beta_{10}(t) = \beta_{10}(t) + \dfrac{\int \beta_{1U}\epsilon \exp\{-(\beta_{0U}+\beta_{1U})^T\epsilon t\} f(\epsilon)d\epsilon }{\int \exp\{-(\beta_{0U}+\beta_{1U})^T\epsilon t\} f(\epsilon)d\epsilon}.$$

\subsection*{Proof of Result~\ref{rslt:cif}}
\begin{proof}
    Firstly, we have 
\begin{align*}
    &P(T\geq t \mid A, U, Z, X)\\
    =& \exp\{-B_{00}(t) - B_{10}(t) - \beta_A^TAt - (\beta_{0X}+\beta_{1X})^TXt - (\beta_{0U}+\beta_{1U})^T\gamma(A, X, Z)t\}\exp\{-(\beta_{0U}+\beta_{1U})^T\epsilon t\}
\end{align*}
And therefore
\begin{align*} S(t\mid A, X, Z) &\equiv P(T\geq t \mid A, Z, X)\\
    &= \exp\{-B_{00}(t) - B_{10}(t) -\Omega(t)- \beta_A^TAt - (\beta_{0X}+\beta_{1X})^TXt - (\beta_{0U}+\beta_{1U})^T\gamma(A, X, Z)t\}\end{align*}
    where we set $\exp\{-\Omega(t)\}=E[\exp\{-(\beta_{0U}+\beta_{1U})^T\epsilon t\}].$

    \begin{align*}
        &E[dN_0^{(a)}\mid T^{(a)}\geq t]\\
        &= E[E\{dN_0^{(a)}\mid T^{(a)}\geq t, U, X, Z\}\mid T^{(a)}\geq t] \\
        &=  E[E\{dN_0\mid T\geq t, A = a, U, X, Z\}\mid T^{(a)}\geq t] \\
        &= E[\beta_{00}(t) + \beta_A^T a + \beta_{0X}^TX + \beta_{0U} U\mid T^{(a)}\geq t]\\
        &= \beta_{00}(t) + \beta_A^T a + E[\beta_{0X}^TX + \beta_{0U} U \mid T^{(a)}\geq t]\\
        &= \beta_{00}(t) + \beta_A^T a + \dfrac{E[\{\beta_{0X}^TX + \beta_{0U} U\}S(t\mid A=a, Z, X, U)]}{P(T^{(a)}\geq t)}\\
        &= \beta_{00}(t) + \beta_A^T a + \dfrac{E[\{\beta_{0X}^TX + \beta_{0U} \gamma(A, Z, X) + \beta_{0U}\epsilon\}S(t\mid A=a, Z, X, U)]}{P(T^{(a)}\geq t)}
    \end{align*}

    Now that we have
   \begin{align*} 
   &P(T^{(a)}\geq t)\\
 &= E\{ S(t\mid A=a, U, Z, X)\} \\
 &= E[\exp\{-B_{00}(t) - B_{10}(t) - \beta_A^Tat - (\beta_{0X}+\beta_{1X})^TXt - (\beta_{0U}+\beta_{1U})^T\gamma(A, X, Z)t\}\exp\{-(\beta_{0U}+\beta_{1U})^T\epsilon t\}]\\
 &= E[\exp(- \beta_A^Tat+ \beta_A^TAt )S(t\mid A, U, Z, X)]\\
 &= E[\exp(- \beta_A^Tat + \beta_A^TAt )S(t\mid A, Z, X)]
\end{align*}
And similarly
\begin{align*}
    & E[\{\beta_{0X}^TX + \beta_{0U} \gamma(A, Z, X) \}S(t\mid A=a, Z, X, U)]\\
    =& E[\{\beta_{0X}^TX + \beta_{0U} \gamma(A, Z, X) \} \exp(- \beta_A^Tat + \beta_A^TAt )S(t\mid A, Z, X, U)]\\
    =& E[\{\beta_{0X}^TX + \beta_{0U} \gamma(A, Z, X) \} \exp(- \beta_A^Tat + \beta_A^TAt )S(t\mid A, Z, X)].
\end{align*}

Finally,
\begin{align*}
    &E[\beta_{0U}\epsilon S(t\mid A=a, Z, X, U)]\\
    =&E[\exp\{-B_{00}(t) - B_{10}(t) - \beta_A^Tat - (\beta_{0X}+\beta_{1X})^TXt - (\beta_{0U}+\beta_{1U})^T\gamma(A, X, Z)t\}\beta_{0U}\epsilon\exp\{-(\beta_{0U}+\beta_{1U})^T\epsilon t\}]\\
    =& E[\exp\{-B_{00}(t) - B_{10}(t) - \beta_A^Tat - (\beta_{0X}+\beta_{1X})^TXt - (\beta_{0U}+\beta_{1U})^T\gamma(A, X, Z)t\}\\&\qquad E\{\beta_{0U}\epsilon\exp\{-(\beta_{0U}+\beta_{1U})^T\epsilon t\}\mid A, Z, X\}]\\
    =& E[\exp\{-B_{00}(t) - B_{10}(t) - \beta_A^Tat - (\beta_{0X}+\beta_{1X})^TXt - (\beta_{0U}+\beta_{1U})^T\gamma(A, X, Z)t\}\times \\
    &\qquad \{\overline\beta_{00}(t) - \beta_{00}(t)\}E\{\exp\{-(\beta_{0U}+\beta_{1U})^T\epsilon t\}\mid A, Z, X\}]\\
    =& \{\overline\beta_{00}(t) - \beta_{00}(t)\} E[\exp\{-B_{00}(t) - B_{10}(t) - \beta_A^Tat - (\beta_{0X}+\beta_{1X})^TXt - (\beta_{0U}+\beta_{1U})^T\gamma(A, X, Z)t\}\\&\qquad \exp\{-(\beta_{0U}+\beta_{1U})^T\epsilon t\}]\\
    =& \{\overline\beta_{00}(t) - \beta_{00}(t)\} E\{S(t\mid A=a, U, Z, X)\}\\
    =& \{\overline\beta_{00}(t) - \beta_{00}(t)\} E\{\exp(- \beta_A^Tat + \beta_A^TAt ) S(t\mid A, U, Z, X)\}\\
    =& \{\overline\beta_{00}(t) - \beta_{00}(t)\} E\{\exp(- \beta_A^Tat + \beta_A^TAt ) S(t\mid A,  Z, X)\}
\end{align*}

We conclude that 
\begin{align*}
    &\lambda_{T_0^{(a)}}(t)\\ &= E[dN_0^{(a)}\mid T^{(a)}\geq t] \\
    &= \overline\beta_{00}(t) +\beta_A^Ta +\\&
    \dfrac{E[\{\beta_{0X}^TX + \beta_{0U} \gamma(A, Z, X) \}\exp\{- \beta_A^Tat + \beta_A^TAt \}S(t\mid A, Z, X)]}{E[\exp\{- \beta_A^Tat  + \beta_A^TAt \}S(t\mid A, Z, X)]}\\
    &= \dfrac{E[\exp\{ \beta_A^TAt \}\{\lambda_0(t\mid A, Z, X)-\beta_A^TA + \beta_A^Ta\}S(t\mid A, Z, X)]}{E[\exp\{  \beta_A^TAt \}S(t\mid A, Z, X)]}\\
\end{align*}

And similarly,
\begin{align*}
    &\lambda_{T_1^{(a)}}(t) \\&= E[dN_1^{(a)}\mid T^{(a)}\geq t] \\
    &= \overline\beta_{10}(t) +\\&
    \dfrac{E[\{\beta_{1X}^TX + \beta_{1U} \gamma(A, Z, X) \}\exp\{- \beta_A^Tat + \beta_A^TAt \}S(t\mid A, Z, X)]}{E[\exp\{- \beta_A^Tat  + \beta_A^TAt \}S(t\mid A, Z, X)]}\\
    &= \dfrac{E[\exp\{\beta_A^TAt \}\lambda_1(t\mid A, Z, X)S(t\mid A, Z, X)]}{E[\exp\{ \beta_A^TAt \}S(t\mid A, Z, X)]}\\
\end{align*}
\end{proof}

\subsection*{Proof of Result~\ref{rslt:cmprisk-nc-period}}

 By Assumptions~\ref{assump:cprisk-ah} and \ref{assump:nc-period-ah}, we have
\begin{align*}
    P(T\geq t\mid A, U, Z, X) &= \exp[-\{B_0(t) + B_A(t)^TA + B_X(t)^TX + B_U(t)^TE(U\mid A, X, Z)\}]\exp\{-B_U(t)^T\epsilon\}
\end{align*}
where $B_0(t) = \int_0^t\{\beta_{00}(s) + \beta_{10}(s)\}ds$, $B_A(t) = \int_0^t\{\beta_{A} + \beta_{1A}(s)\}ds$, $B_X(t) = \int_0^t\{\beta_{0X} + \beta_{1X}(s)\}ds$, $B_U(t) = \int_0^t\{\beta_{0U} + \beta_{1U}(s)\}ds$.

Therefore
\begin{align*}
    P(T\geq t\mid A, Z, X) &= \exp[-\{\overline B_0(t) + B_A(t)^TA + B_X(t)^TX + B_U(t)^TE(U\mid A, X, Z)\}]
\end{align*}

where $\overline B_0(t) = B_0(t) - \log E[\exp\{-B_U(t)^T\epsilon \}]$

Equation~\eqref{eq:cshaz-cmprsk} can also be written as
\begin{align*}
    E\{dN_1(t)\mid A, U, X, Z, T\geq t\} &= \{\beta_{10}(t) + \beta_{1A}(t)^TA + \beta_{1X}(t)^TX + \beta_{1U}(t) U\}dt\\
    &= \{\beta_{10}(t) + \beta_{1A}(t)^TA + \beta_{1X}(t)^TX + \beta_{1U}(t) \gamma(A, X, Z)\}dt + \beta_{1U}(t)\epsilon dt,
\end{align*}

and so
\begin{align*}
    &E\{dN_1(t)\mid A, X, Z, T\geq t\}\\
    =&  \{\beta_{10}(t) + \beta_{1A}(t)^TA + \beta_{1X}(t)^TX + \beta_{1U}(t) \gamma(A, X, Z)\}dt + \beta_{1U}(t)E(\epsilon\mid A, X, Z, T\geq t) dt\\
    &= \{\beta_{10}(t) + \beta_{1A}(t)^TA + \beta_{1X}(t)^TX + \beta_{1U}(t) \gamma(A, X, Z)\}dt + \beta_{1U}(t)\dfrac{E\{\epsilon P(T\geq t\mid A, U, X, Z)\mid A, X, Z\}}{P(T\geq t\mid A, X, Z)}\} dt\\
    &= \{\beta_{10}(t) + \beta_{1A}(t)^TA + \beta_{1X}(t)^TX + \beta_{1U}(t) \gamma(A, X, Z)\}dt + \beta_{1U}(t)\dfrac{E\{\epsilon \exp(-B_U(t)^T\epsilon)\}}{E\{\exp(-B_U(t)^T\epsilon)\}}\} dt\\
    &= \{\widetilde\beta_{10}(t) + \beta_{1A}(t)^TA + \beta_{1X}(t)^TX + \beta_{1U}(t) \gamma(A, X, Z)\}dt
\end{align*}
where $$\widetilde \beta_{10}(t)=\beta_{10}(t) +\beta_{1U}(t)\dfrac{E\{\epsilon \exp(-B_U(t)^T\epsilon)\}}{E\{\exp(-B_U(t)^T\epsilon)\}}.$$

Therefore, the conditional cause-specific hazard function $\lambda_1(t\mid A, X, Z)$ follows an additive hazard regression model
\begin{align*}\lambda_1(t\mid A, X, Z) &= \widetilde\beta_{10}(t) + \beta_{1A}(t)^TA + \beta_{1X}(t)^TX + \beta_{1U}(t) \gamma(A, X, Z)\\
&= \begin{cases} \widetilde\beta_{10}(t) + b_A^TA + b_{X1}^TX + b_{U1}\gamma(A, X, Z) & \text{if $t\leq D$} \\
\widetilde\beta_{10}(t) + b_{X2}^TX + b_{U2}\gamma(A, X, Z) & \text{if $t>D$}\end{cases}\end{align*}

Similarly, \begin{align*}
\lambda_0(t\mid A, X, Z) &= \widetilde\beta_{00}(t) + \beta_{A}^TA + \beta_{0X}^TX + \beta_{0U}^T \gamma(A, X, Z) \end{align*}
where $$\widetilde \beta_{00}(t)=\beta_{00}(t) +\beta_{0U}\dfrac{E\{\epsilon \exp(-B_U(t)^T\epsilon)\}}{E\{\exp(-B_U(t)^T\epsilon)\}}.$$

Under Assumption~\ref{assump:u-lm0}, we have
\begin{equation}
        \lambda_1(t\mid A, X, Z) = b_{12}^*(t) + (b_{A2}^*)^T A + (b_{X2}^*)^T X + (b_{Z2}^*)^T Z, \quad t>D\label{eq:first-stage}
\end{equation}

where $b_{12}^*(t) = \widetilde\beta_{10}(t) + b_{U2}\gamma_0$, $b_{A2}^* = b_{U2}\gamma_A$, $b_{X2}^*=b_{X2}+b_{U2}\gamma_X$, $b_{Z2}^* = b_{U2}\gamma_Z$.

Let $\mu(A, X, Z) = (b_{A2}^*)^T A + (b_{X2}^*)^T X + (b_{Z2}^*)^T Z$. We have that 

\begin{equation}\label{eq:anx-model}
    \lambda_1(t\mid A, X, Z) = b_{11}^*(t) + (b_{A1}^*)^T A + (b_{X1}^*)^T X + b_{U1}^*\mu(A, X, Z), \quad t\leq D
\end{equation}
where $b_{11}^*(t) = \widetilde\beta_{11}(t)$, $b_{A1}^*=b_A$, $b_{X1}^*=b_{X1}-b_{U1}b_{X2}/b_{U2}$ and $b_{U1}^*=b_{U1}/b_{U2}$. We also have 
\begin{equation}\label{eq:second-stage}
\lambda_0(t\mid A, X, Z) = \beta_{00}^*(t) + \beta_{A}^TA + (\beta_{0X}^*)^TX + \beta_{0U}^* \mu(A, X, Z) \end{equation}
where $\beta_{00}^*(t) = \widetilde \beta_{00}(t)$, $\beta_{0X}^* = \beta_{0X}-\beta_{0U}b_{X2}/b_{U2}$ and $\beta_{0U}^*=\beta_{0U}/b_{U2}$.

To estimate $\beta_A$, we may use the following two-stage-least-squares approach:

\begin{enumerate}
    \item Fit the additive hazard regression model for the cause-specific hazard according to Eq.~\eqref{eq:first-stage}, using only subjects with censored event times $\widetilde T>D$. Obtain the parameter estimates $\widehat b_{A2}^*$, $\widehat b_{X2}^*$ and $\widehat b_{Z2}^*$.
    \item Let $\widehat \mu(A, X, Z)=(\widehat b_{A2}^*)^T A + (\widehat b_{X2}^*)^T X + (\widehat b_{Z2}^*)^T Z$.
    \item Fit the additive hazard regression model for the cause-specific hazard according to Eq.~\eqref{eq:second-stage} with $\mu(A,X,Z)$ replaced with $\widehat \mu(A, X, Z)$, using all subjects. The parameter of interest is the regression coefficient of $A$, denoted as $\widehat\beta_A$. Inference for $\widehat\beta_A$ may be obtained via nonparametric bootstrap.
\end{enumerate}

We continue to derive $P(T^a>t)$. We have that
\begin{align*}
    P(T^a > t) &= E\{P(T^a > t\mid U, Z, X)\}\\
    &= E\{P(T^a > t\mid A=a, U, Z, X)\}\qquad\mbox{(Exchangeability)}\\
    &= E\{P(T> t\mid A=a, U, Z, X)\}\qquad\mbox{(Consistency)}\\
    &=  E\left\{\exp[-\{B_0(t) + B_A(t)^Ta + B_X(t)^TX + B_U(t)^TU\}]\right\}\\
    &= E\bigg[\exp\{-B_A(t)^Ta + B_A(t)^TA \}P(T> t\mid A, U, Z, X)\bigg]\\
    &=E\bigg[\exp\{-B_A(t)^Ta + B_A(t)^TA \}E\{P(T> t\mid A, U, Z, X)\mid A, X, Z\}\bigg] \\
    &= E\bigg[\exp\{-B_A(t)^Ta + B_A(t)^TA \}P(T> t\mid A, Z, X)\bigg] 
\end{align*}

We also have
\begin{align*}
    E\{dN_0^a(t)\mid T^a \geq t\} &= E[E\{dN_0^a(t)\mid U, X, Z, T^a \geq t\}\mid T^a \geq t] \\
    &= E[E\{dN_0^a(t)\mid A = a, U, X, Z, T^a \geq t\}\mid T^a \geq t] \\
    &= E[E\{dN_0(t)\mid A = a, U, X, Z, T \geq t\}\mid T^a \geq t]\\
    &= E\{ \beta_{00}(t) + \beta_A^T a + \beta_{0X}^TX + \beta_{0U}U\mid T^a \geq t]dt\\
    &= \left\{\beta_{00}(t) + \beta_A^T a + E\{\beta_{0X}^TX + \beta_{0U}U \mid T^a \geq t]\right\}dt\\
    &= \left\{\beta_{00}(t) + \beta_A^T a + \dfrac{E[\{\beta_{0X}^TX + \beta_{0U}U\}P(T\geq t\mid A=a, X, Z, U)]}{P(T^a\geq t)}\right\}dt\\
    &= \left\{\beta_{00}(t) + \beta_A^T a + \dfrac{E[\{\beta_{0X}^TX + \beta_{0U}\gamma(A, Z, X) + \beta_{0U}\epsilon\}P(T\geq t\mid A=a, X, Z, U)]}{P(T^a\geq t)}\right\}dt
\end{align*}

Now \begin{align*}
    &E[\{\beta_{0X}^TX + \beta_{0U}\gamma(A, Z, X)\}P(T\geq t\mid A=a, X, Z, U)]\\
    =&E[\{\beta_{0X}^TX + \beta_{0U}\gamma(A, Z, X)\}\exp\{-B_A(t)^Ta + B_A(t)^TA \}P(T\geq t\mid A, Z, X, U)]\\
    =&E[\{\beta_{0X}^TX + \beta_{0U}\gamma(A, Z, X)\}\exp\{-B_A(t)^Ta + B_A(t)^TA \}E\{P(T\geq t\mid A, Z, X, U)\mid A, Z, X\}]\\
    =& E[\{\beta_{0X}^TX + \beta_{0U}\gamma(A, Z, X)\}\exp\{-B_A(t)^Ta + B_A(t)^TA \}P(T\geq t\mid A, Z, X)]
\end{align*}
and
\begin{align*}
    &E[\beta_{0U}\epsilon P(T\geq t\mid A=a, X, Z, U)]\\
    &= E[\beta_{0U}\epsilon \exp[-\{B_0(t) + B_A(t)^Ta + B_X(t)^TX + B_U(t)^T\gamma(A, Z, X)\}]\exp\{-B_U(t)^T\epsilon\}] \\
    &=E\left[E\{\beta_{0U}\epsilon\exp\{-B_U(t)^T\epsilon\}\mid A, Z, X\} \exp[-\{B_0(t) + B_A(t)^Ta + B_X(t)^TX + B_U(t)^T\gamma(A, Z, X)\}]\right]\\
    &= E\left[\{\widetilde \beta_{00}(t)-\beta_{00}(t)\}\exp[-\{B_0(t) + B_A(t)^Ta + B_X(t)^TX + B_U(t)^T\gamma(A, Z, X)\}]E[\exp\{-B_U(t)^T\epsilon\}]\right]\\
    &= \{\widetilde \beta_{00}(t)-\beta_{00}(t)\}E[\exp\{-B_A(t)^Ta + B_A(t)^TA \}P(T\geq t\mid A, Z, X)].
\end{align*}

We obtain that
\begin{align*}
     E\{dN_0^a(t)\mid T^a \geq t\} &= \left[\widetilde \beta_{00}(t) + \beta_A^Ta + \dfrac{E[\{\beta_{0X}^TX + \beta_{0U}\gamma(A, Z, X)\}\exp\{-B_A(t)^Ta + B_A(t)^TA \}P(T\geq t\mid A, Z, X)]}{E[\exp\{-B_A(t)^Ta + B_A(t)^TA \}P(T\geq t\mid A, Z, X)]}\right]dt
\end{align*}

We conclude that 
\begin{align*}\lambda^a_0(t)& = \widetilde \beta_{00}(t) + \beta_A^Ta + \dfrac{E[\{\beta_{0X}^TX + \beta_{0U}\gamma(A, Z, X)\}\exp\{-B_A(t)^Ta + B_A(t)^TA \}P(T\geq t\mid A, Z, X)]}{E[\exp\{-B_A(t)^Ta + B_A(t)^TA \}P(T\geq t\mid A, Z, X)]}\\
&=\dfrac{E[\{\lambda_0(t\mid A, Z, X)+\beta_A^Ta -\beta_A^TA \}\exp\{-B_A(t)^Ta + B_A(t)^TA \}P(T\geq t\mid A, Z, X)]}{E[\exp\{-B_A(t)^Ta + B_A(t)^TA \}P(T\geq t\mid A, Z, X)]}
\end{align*}

\newpage
\section{Multiple NCOs, potentially different data types}
Suppose $U$ is a multi-dimensional vector of unmeasured confounders. We modify Assumptions 1 as follows:

\begin{assumption}[Additive hazards model]\label{assump:ah-multi-nco}
    The conditional hazard function for the outcome $T$ given $A, U, X$, $\lambda_T(t\mid A, U, X)$, satisfies
    \begin{equation}\label{eq:ah}
        \lambda_T(t\mid A, U, X)=\beta_0(t) + \beta_A^TA + \beta_X^TX + \beta_U^TU. 
    \end{equation}

\end{assumption}

Suppose Assumptions 2-5 still hold and we have multiple NCOs, potentially with different data types. Let $W=(W_{11},\dots,W_{1K_1},W_{21},\dots,W_{2K_2},W_{31},\dots,W_{3K_3})$ be the vector of NCOs, where the distribution of $W_{11}$, \dots, $W_{1K_1}$ each follows a linear model, the distribution of $W_{21}$, \dots, $W_{2K_2}$ each follows a GLM with log link, and  $W_{31}$, \dots, $W_{3K_3}$ are time-to-event variables each of each follows an additive hazards model. Formally:

\begin{assumption}[Model assumptions for multiple NCOs]\label{assum:multi-nco}
    \begin{align*}
        E(W_{1j}\mid A, U, X) &= c_{01j} + c_{U1j}^TU + c_{X1}^TX&&\mbox{for $j=1,\dots, K_1$};\\
        E(W_{2j}\mid A, U, X) &= \exp\{c_{02j} + c_{02j}^TU + c_{X2j}^TX\}&&\mbox{for $j=1,\dots, K_2$};\\
        \lambda_{W_{3j}}(t\mid A, U, X) &= c_{03j}(t) + c_{U3j}^TU + c_{X3j}^TX&&\mbox{for $j=1,\dots, K_3$}.
    \end{align*}
\end{assumption}

With derivation similar to those in Sections 2.1-2.3, we can show that 

\begin{align}
        E(W_{1j}\mid A, Z, X) &= c_{01j}^* + (c_{A1j}^*)^TA + (c_{Z1j}^*)^TZ + (c_{X11}^*)^TX&&\mbox{for $j=1,\dots, K_1$};\label{eq:mnco-lin}\\
        E(W_{2j}\mid A, Z, X) &= \exp\{c_{02j}^* + (c_{A2j}^*)^TA + (c_{Z2j}^*)^TZ + (c_{X2j}^*)^TX\}&&\mbox{for $j=1,\dots, K_2$};\label{eq:mnco-loglin}\\
        \lambda_{W_{3j}}(t\mid A, Z, X) &= c_{03j}^*(t) + (c_{A3j}^*)^TA + (c_{Z3j}^*)^TZ + (c_{X3j}^*)^TX&&\mbox{for $j=1,\dots, K_3$}.\label{eq:mnco-ah}
    \end{align}
where  $c_{01j}^*=c_{0j1} +c_{U1j}^T\gamma_{0}$, $c_{A1j}^*=\gamma_{A}^Tc_{U1j}$, $c_{Z1j}^*=\gamma_Z^Tc_{U1j}$, $c_{X1j}^*=\gamma_U^Tc_{U1j} + c_{X1j}$, $c_{02j}^*=c_{02j} + \log\{\int \exp(c_{U2j}^Te)dF_\epsilon(e)\} + c_{U2j}^T\gamma_{0}$, $c_{A2j}^*=\gamma_{A}^Tc_{U2j}$, $c_{Z2j}^*=\gamma_Z^Tc_{U2j}$, $c_{X2j}^*=\gamma_U^Tc_{U2j} + c_{X2j}$, $c_{03j}^*(t)= c_{03j}(t) - \partial[\log\int\exp\{-tc_{U3j}^Te\}dF_\epsilon(e)]/\partial t + c_{U3j}^T\gamma_{0}$, $c_{A3j}^*=\gamma_{A}^Tc_{U3j}$, $c_{Z3j}^*=\gamma_Z^Tc_{U3j}$, and $c_{X3j}^*=\gamma_U^Tc_{U3j} + c_{X3j}$.

We write 
\begin{align*}
&\mathbf \mu(A,Z,X) = \begin{pmatrix}c_{011}^* + (c_{A11}^*)^TA + (c_{Z11}^*)^TZ + (c_{X11}^*)^TX\\ \vdots \\ c_{01K_1}^* + (c_{A1K_1}^*)^TA + (c_{Z1K_1}^*)^TZ + (c_{X1K_1}^*)^TX \\c_{021}^* + (c_{A21}^*)^TA + (c_{Z21}^*)^TZ + (c_{X21}^*)^TX\\\vdots \\ c_{02K_2}^* + (c_{A2K_2}^*)^TA + (c_{Z2K_2}^*)^TZ + (c_{X2K_2}^*)^TX\\(c_{A31}^*)^TA + (c_{Z31}^*)^TZ + (c_{X31}^*)^TX\\\vdots \\(c_{A3K_3}^*)^TA + (c_{Z3K_3}^*)^TZ + (c_{X3K_3}^*)^TX\end{pmatrix},\\
&\mathbf c_0=\begin{pmatrix}c_{011}\\\vdots\\c_{01K_1}\\c_{021} + \log\{\int \exp(c_{U21}^Te)dF_\epsilon(e)\}\\\vdots \\ c_{02K_2} + \log\{\int \exp(c_{U2K_2}^Te)dF_\epsilon(e)\}\\ -c_{U31}^T\gamma_0\\\vdots \\-c_{U3K_3}^T\gamma_0 \end{pmatrix},\\
&\mathbf c_U =\begin{pmatrix}c_{U11} & c_{U12}&\dots &c_{U1K_1}&c_{U21}&\dots &c_{U2K_2}&c_{U31}&\dots &c_{U3K_3}
\end{pmatrix},\\
&\mathbf c_X =\begin{pmatrix}c_{X11} & c_{X12}&\dots &c_{X1K_1}&c_{X21}&\dots &c_{X2K_2}&c_{X31}&\dots &c_{X3K_3}
\end{pmatrix}.
\end{align*}

We have that 
\begin{equation}\label{eq:lin-predictor}\mu(A,Z,X) = \mathbf c_0 + \mathbf c_U^TE(U\mid A, Z, X) + \mathbf c_X^TX.\end{equation}

By Equations (1) and \eqref{eq:lin-predictor}, we have
\begin{equation}\label{eq:2s-multi-nco}
\lambda_T(t\mid A, Z, X) =  \beta_{04}^{*}(t)+\beta_A^TA + (\beta_{U1}^*)^TE(W\mid A, Z, X) + (\beta_{X1}^*)^T(X)
\end{equation} where $\beta_{04}^*(t) = \widetilde\beta_0(t) - \beta_U^T(\mathbf c_{U}^T)^+\mathbf c_{0}$, $\beta_{U4}^* = \mathbf c_{U}^+\beta_U$, and $\beta_{X4}^*=\beta_X - \mathbf c_{X}\mathbf c_{U}^+\beta_U$. Here $M^+$ denotes the left-inverse of a matrix $M$, if it exists. Equations \eqref{eq:mnco-lin}-\eqref{eq:mnco-ah} and \eqref{eq:2s-multi-nco} justify the two-stage regression approach in Algorithm 4.

Under Assumption~\ref{assum:multi-nco}, the U-relevance assumption akin to Assumptions 7A-7C is:

\begin{assumption}[U-relevance]\label{assump:completeness-multi-nco}
    The matrix $\mathbf c_U$ has full row rank and $\gamma_Z\neq 0$.
\end{assumption}

We see that Assumption~\ref{assump:completeness-multi-nco} is more likely to hold with more NCEs and NCOs that are associated with $U$. This matches the intuition that with more negative control variables, we have more information for unmeasured confounding bias adjustment.

\newpage
\section{Simulation setting}

In Section 5, we simulate the data based on the following data generating mechanism:

\begin{align*}
        U, X &\sim Uniform(0,1)\\
        A\mid U, X&\sim Bernoulli(\dfrac{1}{1 + \exp(-3 + 5U + X)})\\
        T\mid A, U, X &\sim Exponential(0.2 + 0.2 A +\beta_{UY}U + 0.2 X)\\
        C &= 5\\
        W\mid U, X&\sim N\left(\begin{pmatrix}
            0.5c_U U + 0.2X\\
            2c_U U + X
        \end{pmatrix},\begin{pmatrix}
            0.1^2 & 0 \\
            0 & 0.25^2
        \end{pmatrix}\right)\\
        Z\mid U, X&\sim N\left(\begin{pmatrix}
            c_U U + 0.5X\\
            0.5c_U U + 2X
        \end{pmatrix},\begin{pmatrix}
            0.5^2 & 0 \\
            0 & 0.2^2
        \end{pmatrix}\right)
    \end{align*}

\newpage
\section{Summary of assumptions of the proximal two-stage regression (P2SR-Surv) and their plausibility in the RHC data example}

\begin{table}[!htbp]
    \centering
        \caption{Summary of assumptions of P2SR-Surv and their plausibility in the RHC data example}
    \label{tab:assumpts}
    \begin{tabular}{|l|l|}
    \hhline{|=|=|}
     Assumption    & Plausibility in the RHC example \\
    \hhline{|=|=|}
    \begin{minipage}{0.4\textwidth}
    \textbf{Additive hazards model:} The conditional hazard function given $(A, U, X)$ is 
    
    $\lambda_T(t\mid A, U, X)=$
    
    $\qquad \beta_{0}(t) + \beta_A^T A + \beta_{X}^TX + \beta_{U}^T U.$
\end{minipage}     &  \begin{minipage}{0.6\textwidth}The assumption holds if the conditional hazard function of death after hospital admission is additive and linear in the considered covariates and the unmeasured confounder, and that the effects are constant during the follow-up.

The assumption is violated if
\begin{enumerate}
    \item the conditional hazard function of death is non-additive or nonlinear in the variables;
    \item the effects are time-constant.
\end{enumerate}
Time-varying effects are more likely to occur during long follow-up. In our analysis, we only consider death up to 180 days after hospital admission so that the follow-up time is moderate.

\end{minipage}\\
\hline \begin{minipage}{0.4\textwidth}
\textbf{Negative control variables}:  The NCE $Z$ and NCO $W$ satisfy $A\indep W\mid U, X$ and $Z\indep (W, T)\mid A, U, X$.
\end{minipage} & \begin{minipage}{0.6\textwidth} We select  $\text{PaO}_2/\text{FiO}_2$ and $\text{PaCO}_2$ as the NCE $Z$, the blood pH and hematocrit as the NCO $W$. This assumption is reasonable in the RHC data since (1) no known biological mechanism indicates a direct effect of RHC on  blood pH and hematocrit,  (2) no known biological mechanism indicates a direct effect of $\text{PaO}_2/\text{FiO}_2$ and $\text{PaCO}_2$ on mortality, and (3) no known biological mechanism indicates a direct effect of $\text{PaO}_2/\text{FiO}_2$ and $\text{PaCO}_2$ on  blood pH and hematocrit. Any associations between the above three variable pairs are likely an result of the underlying confounding by patients' disease severity.

The negative control assumption may be violated if RHC or a change in $\text{PaO}_2/\text{FiO}_2$ or $\text{PaCO}_2$ may  cause a direct change in blood pH or hematocrit, or if a change in $\text{PaO}_2/\text{FiO}_2$ or $\text{PaCO}_2$ would directly increase or decrease a patient's mortality.

\end{minipage}\\
\hline \begin{minipage}{0.4\textwidth}
    \textbf{Conditionally independent censoring: }$C\indep T\mid A, X, Z$
\end{minipage} & \begin{minipage}{0.6\textwidth}This assumption holds if conditioning on $(A, X, Z)$, loss-of-follow-up is uncorrelated with a patient's mortality. The assumption may not hold if loss of follow-up is induced by an unmeasured confounder beyond observed covariates, such that in some strata defined by $(A, X, Z)$, patients who were censored were at a higher or lower mortality than those who weren't. \end{minipage}\\
\hline
\begin{minipage}{0.4\textwidth}
    \textbf{Location-shift model for $U$: }$U=E(U\mid A, Z, X)+\epsilon$ where the distribution of $\epsilon$ is unrestricted other than $E(\epsilon)=0$ and $\epsilon\indep (A, Z, X)$.
\end{minipage} &
\begin{minipage}{0.6\textwidth}
    This strong assumption states that the conditional distribution of $U$ given $(A, Z, X)$ follows a location-shift model, and therefore depends on the latter only through its mean, so that the residual error $\epsilon$ is independent of $(A, Z, X)$. In the RHC example, this assumption requires that the variation of the underlying disease severity is homoscedastic, i.e. approximately the same across strata defined by $A, Z, X$.
\end{minipage}\\
\hline 
\begin{minipage}{0.4\textwidth}
\textbf{Linear mean model for $U$:} 

$E(U\mid A, Z, X)=$

$\gamma_{0}+\gamma_{A}A + \gamma_{Z}^TZ + \gamma_{X}^TX+ \gamma_{ZA}^TZA + \gamma_{XA}^TXA.$

\end{minipage} &

\begin{minipage}{0.6\textwidth}This strong assumption requires that the distribution of the unmeasured confounder (disease severity) follows a linear mean model conditioning on $(A,X,Z)$ as stated. In our analysis, we include the interaction terms $\gamma_{ZA}^TZA$ and $\gamma_{XA}^TXA$ to allow more flexible mean model of $U$ compared with Assumption 5.
    
\end{minipage}\\

    \hhline{|=|=|}
    \end{tabular}

\end{table}

\begin{table}[!htbp]
    \centering
        \caption{Summary of assumptions of P2SR-Surv and their plausibility in the RHC data example (continued)}
    \label{tab:assumpts}
    \begin{tabular}{|l|l|}
    \hhline{|=|=|}
     Assumption    & Plausibility in the RHC example \\
    \hhline{|=|=|}
\begin{minipage}{0.4\textwidth}
    \textbf{NCO follows a linear mean model: }

    $E(W\mid A, U, X)=c_{0} + c_{U}^TU+c_{X}^TX$
\end{minipage} & \begin{minipage}{0.6\textwidth}This assumption requires a linear mean model of the NCOs (blood PH and hematocrit) conditioning on $(A, U,X)$. The assumption may not hold if  the assumed mean model is incorrect for either NCO.
    
\end{minipage}
    \\
    \hline
    \begin{minipage}{0.4\textwidth}
      (1) $c_{U}$ has full row rank and (2) $\gamma_{ZA}\neq 0$ or $\gamma_Z\neq 0$. 
    \end{minipage}
    & \begin{minipage}{0.6\textwidth}This assumption requires that both $W$ and $Z$ are correlated with the unmeasured confounder $U$. In the RHC example, the selected negative control variables are all indicators of disease severity, so this assumption is likely to hold.\end{minipage}\\
    \hhline{|=|=|}
\end{tabular}
\end{table}

\printbibliography[heading=subbibliography]

\end{refsection}

\end{document}